\documentclass[aps, superscriptaddress,amsmath,amssymb,floatfix, twocolumn, notitlepage, bm,braket]{revtex4-1}
\usepackage{times}
\usepackage{graphics}
\usepackage{graphicx}
\usepackage{color}
\usepackage{bm}
\usepackage{braket}
\usepackage{mathtools}
\usepackage{mathcomp}
\usepackage[caption=false,justification=raggedright,position=top,singlelinecheck=off]{subfig}
\usepackage{soul}
\usepackage{tabularx}
\usepackage{calrsfs}
\usepackage[vcentermath]{youngtab}
\usepackage[mathscr]{euscript}
\usepackage[normalem]{ulem}

\newcommand{\dn}{\downarrow}

\newcommand{\up}{\uparrow}

\newcommand{\enni}{\noindent}
\newcommand{\enbe}{\begin{equation}}
\newcommand{\enee}{\end{equation}}
\newcommand{\enba}{\begin{align}}
\newcommand{\enea}{\end{align}}

\begin{document}

\title{Matrix-pairing states in the alkaline Fe-selenide superconductors: exotic Josephson junctions}

\author{Emilian M.\ Nica}
\affiliation{Department of Physics
Box 871504
Arizona State University
Tempe, Arizona  85287-1504}
\email[Corresponding author: ]{enica@asu.edu}

\author{Qimiao Si}
\affiliation{Department of Physics and Astronomy, Rice University, 6100 Main St, Houston, TX, 77005, USA}
\affiliation{Rice Center for Quantum Materials, Rice University, 6100 Main St, Houston 77005 TX, USA}

\author{Onur Erten}
\affiliation{Department of Physics
Box 871504
Arizona State University
Tempe, Arizona  85287-1504}

\begin{abstract}
True to their unconventional nature, multi-band alkaline Fe-selenides and, more recently, the heavy-fermion CeCu$_{2}$Si$_{2}$ 
have shown experimental signatures of fully-gapped but sign-changing superconductivity. 
A two-orbital pairing state, called $s\tau_{3}$, with non-trivial matrix structure, was proposed as a candidate able to reconcile 
the seemingly contradictory properties of these superconductors. Motivated by the non-trivial orbital structure of 
the proposed $s\tau_{3}$ state, which has opposite signs for the pairing functions of the two orbitals, 
we study prototypical Josephson junctions where at least one of the leads is in a superconducting state of this kind. 
An analysis of these junctions in the limit of two degenerate orbitals (bands) and with a simple form of junction hybridization reveals several remarkable properties. One is the emergence of gapless, purely electron- and hole-like bound states for $s\tau_{3}-N-s\tau_{3}$ junctions 
  with arbitrary global phase difference between the leads, and likewise for $s\tau_{3}-N-I$ junctions. 
  The other is the absence of static Josephson currents when both leads are superconducting.
   In both of these signatures, $s\tau_{3}$ junctions are dramatically different from more conventional Josephson junctions. 
   We also find that the gapless bound states are protected by an orbital-exchange symmetry, 
   although the protection is not topological. 
   Junctions which break this symmetry, such as $s\tau_{3}-N-s$, have gapped Andreev bound states. 
   In general, the Josephson effect also re-emerges once the degeneracy of the two orbitals is lifted. 
  We support these conclusions via analytical and numerical results for the bound states, 
  together with microscopic calculations of the Josephson current. Our results indicate that junctions involving $s\tau_{3}$ pairing 
  in alkaline Fe-selenides will generically have bound states with a small gap together with a greatly suppressed Josephson current.  
\end{abstract}

\maketitle

\section{Introduction}

Unconventional superconductors (SC's) have led to some of the most important questions in condensed matter physics. 
One such puzzle came to light not too long ago in a branch of the Fe-based family, the alkaline Fe-selenides~\cite{Lee_Science_2017}. 
On one hand these SC's are fully-gapped, as evidenced by ARPES studies~\cite{Mou, Wang_2011, Xu, Wang_2012}, 
while on the other hand, they exhibit an in-gap spin-resonance in inelastic neutron scattering experiments~\cite{Park, Friemel}. 
The first of these features points toward an $s$-wave pairing state, while the second implies a pairing state 
which changes sign under a $\pi/2$ rotations~\cite{Eschrig, Stockert_Nat_Phys_2011,Maier_PRB_2011, Dai_RMP2015,Si2016}, 
such as $d$-wave pairing. These seemingly mutually-exclusive, traits were shown to be reconcilable in a pairing state which transforms 
as a sign-changing $B_{1g}$ representation of the point group, which nonetheless leads to a fully-gapped SC state. 
We called this state $s\tau_{3}$~\cite{Nica_Yu} since it consists of a $s_{x^{2}y^{2}}$ form factor 
multiplied by a $\tau_{3}$ Pauli matrix in a $d_{xz}, d_{yz}$ two-orbital space appropriate to the alkaline Fe-selendides. 
Underlying the remarkable properties of this pairing candidate is the non-trivial $\tau_{3}$ matrix structure in orbital space, 
which ensures that it transforms as $B_{1g}$. In the tight-binding appropriate to the alkaline Fe-selenides this pairing state 
can also be thought of as an effective $d+d$ intra- and inter-\emph{band} pairing~\cite{Nica_Si_npj_2021}. 
$s\tau_{3}$ pairing was stabilized in a realistic five-orbital model of the alkaline Fe-selenides~\cite{Nica_Yu}. 
Remarkably, a similar experimental landscape has also recently emerged in the venerable heavy-fermion CeCu$_{2}$Si$_{2}$, 
the first-discovered unconventional SC~\cite{Steg_1979}. Believed to be a typical $d$-wave for almost it's entire history, 
CeCu$_{2}$Si$_{2}$ was recently found to exhibit a small gap for temperatures well below $T_{c}$ from specific heat~\cite{Kittaka} 
and London penetration depth measurements~\cite{Pang, Yamashita}. In-gap spin resonances in the inelastic neutron spectrum~\cite{Stockert_Nat_Phys_2011} as well as other indicators single out CeCu$_{2}$Si$_{2}$ 
as another example of gapped but sign-changing superconductivity. 
A pairing candidate, which incorporates the orbital and spin-structure of this compound into a non-trivial matrix pairing 
was also recently proposed by two of us~\cite{Nica_Si_npj_2021}. More generally, the $s\tau_{3}$ pairing state represents one of the most dramatic forms of the overall notion 
of orbital-selective superconducting pairing introduced in the context of 111 iron pnictides~\cite{Yu_Zhu_Si},
which has also been subsequently discussed in other iron pnictides~\cite{Yin_Haule_Kotliar,Coleman} and 
the nematic FeSe~\cite{Sprau, HY_Hu2019,Yu_Zhu_Si_2018}.

\begin{figure}[ht!]
\includegraphics[width=1.0\columnwidth]{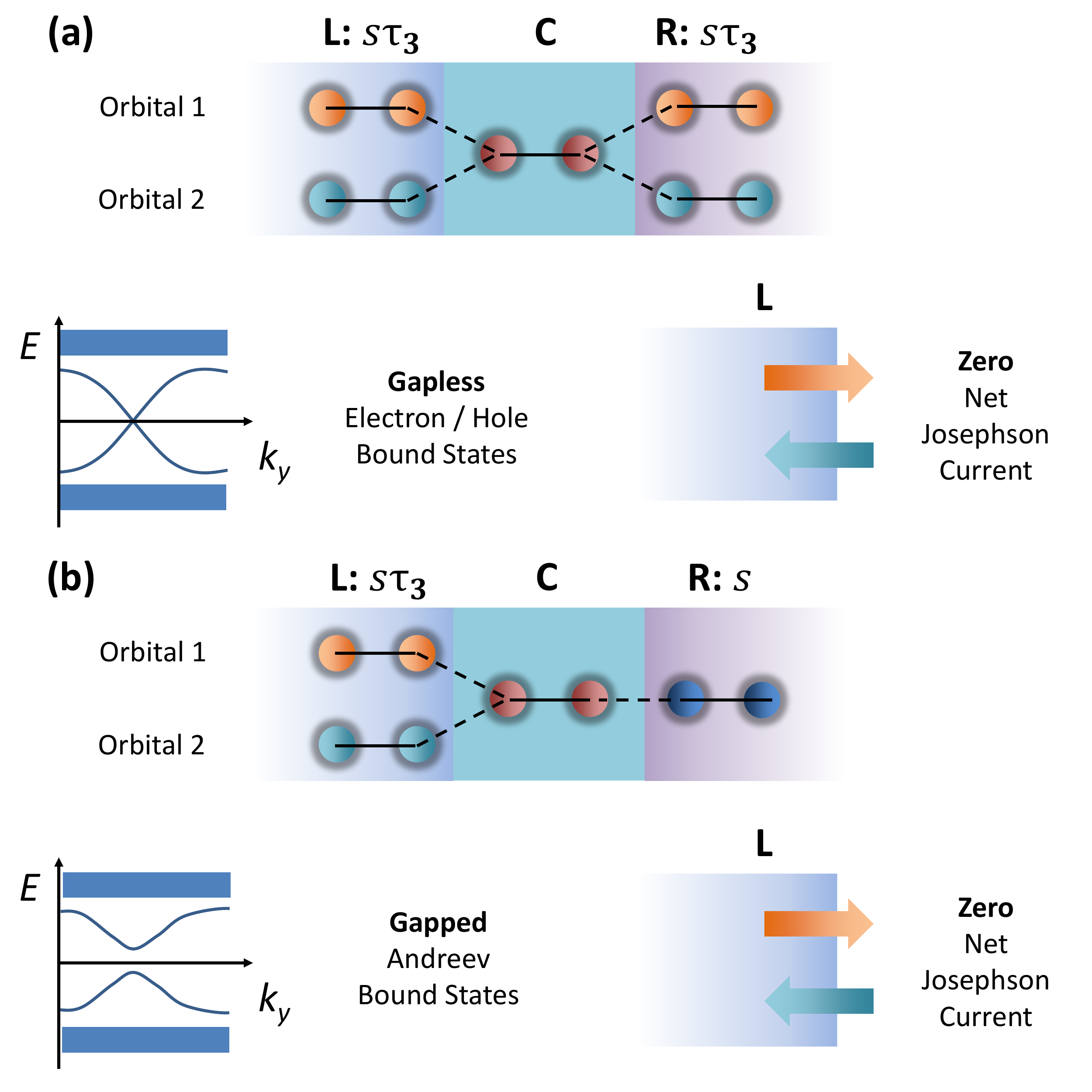}
\caption{Summary of our results for Josephson junctions along $x$ where at least one of the Left (L) and Right (R) leads is in a $s\tau_{3}$ pairing state. In that case, the pairing functions of the two orbitals have opposite signs and thus have non-trivial structure on orbital space. We show the most important results in the limit where the two orbitals are degenerate and couple identically to a single orbital in the Center (C) metallic part.~(a) $s\tau_{3}-N-s\tau_{3}$ junctions where the L and R leads are identical up to a global phase difference. For arbitrary global phase differences between L and R leads, the bound states are either electron- or hole-like and become gapless for a set of conserved momenta $k_{y}$ . In this limit, the static Josephson current is zero, since the contributions of the two orbital sectors effectively cancel. For almost degenerate bands, as expected for systems with $s\tau_{3}$ pairing, the bound states acquire a small gap via Andreev scattering, although exceptions can occur for junctions along $z$. A weak static Josephson current, determined by band-splitting near the FS is also expected. $s\tau_{3}-N-I$ junctions, where the R lead is in an insulating state, has similar bound states.~(b) $s\tau_{3}-N-s$ junction, where the R lead consists of a single orbital in a trivial $s$-wave pairing state. In the same degenerate limit, there are gapped Andreev bound states, but a vanishing Josephson current. For small band splitting, a weak Josephson current is expected.}
\label{Fig:Drft_Schmts_v1}
\end{figure}

In Ref.~\cite{Nica_Si_npj_2021}, we also pointed out that the $s\tau_{3}$ pairing state and a similarly-constructed 
microscopic candidate in CeCu$_{2}$Si$_{2}$ are both equivalent to $d+d$ intra- and inter-band pairing in the band basis. 
This equivalence holds when periodic boundary conditions are imposed along all of the axes. 
Moreover, we discussed the similarities between $d+d$ and $^{3}$He-B, where a similar $p+p$ pairing can be realized in spin space. 
In view of the topologically-protected edge states of $^{3}$He-B, a natural question is 
whether $d+d$ pairing can also have non-trivial edge states. This consideration opens up a new direction in the exploration of both the general $d+d$ pairing and 
actual microscopic states such as $s\tau_{3}$.
For definiteness,  in this work we focus on Josephson junctions as these show features which are specific to 
this type of matrix-pairing and therefore can facilitate its identification by experiments. We reserve a more detailed discussion of the edge state spectrum for future work.

Although some of the bulk properties of pairing candidates with non-trivial matrix structure have already been mentioned, 
much remains to be explored as far as characteristic experimental signatures are concerned. 
One promising route involves Josephson junctions, 
which in principle can take advantage of the inherent phase difference between the two orbital sectors in $s\tau_{3}$ pairing. 
Motivated by this observation, we study prototypical S-N-S and S-N-I junctions 
where at least one of the leads is in a $s\tau_{3}$ pairing state. 
Although our study is specifically geared toward junctions with $s\tau_{3}$ pairing states and thus target the alkaline 
Fe-selenides, we expect that some of the features discussed here can also manifest for similar pairing candidates with non-trivial matrix structure 
 as proposed for CeCu$_{2}$Si$_{2}$~\cite{Nica_Si_npj_2021}.

To isolate the most striking features of this pairing state, we describe the superconducting leads via effective two-orbital models.  
Furthermore, in order to capture the effects of the intrinsic phase difference between the pairing functions of  the two orbitals, 
we consider junctions where  the two orbitals couple to a single orbital in the metallic part. 
As we show, the bound states in these rather unusual setups differ sharply from those in more conventional junctions, 
which ignore cross-coupling between the various orbital.  More precisely, we examine setups where the Left (L) lead, 
Center (C) metallic, and Right (R) leads are arranged in $s\tau_{3}-N-s\tau_{3}$, $s\tau_{3}-N-I$ (insulator), 
and $s\tau-N-s$ (trivial single-channel s-wave SC) junctions. We assume that the coupling to the junction does not destabilize $s\tau_{3}$ pairing. Our salient results are as summarized below.

\subsection{Summary of main results}

Our most striking results occur when the two orbitals of the $s\tau_{3}$ SC leads are degenerate. In this limit, both intra- and inter-orbital hybridization terms which lead to band splitting are absent, and therefore the two orbitals correspond to two degenerate bands. The precise definitions of the intra- and inter-orbital hybridizations are discussed in Sec.~\ref{Sec:Mdls}. In addition, the two orbitals couple identically to a single orbital of the C part. Due to the effects of the non-trivial orbital structure of $s\tau_{3}$ pairing and to the equal coupling to a single C orbital, the usual Andreev scattering processes are effectively ``frustrated'', leading to gapless bound states which are either electron- or hole-like for $s\tau_{3}-N-s\tau_{3}$ for arbitrary global phase differences between L and R leads. This also holds for $s\tau_{3}-N-I$ junctions. The bound states are non-trivial since they decay into the SC leads. We also find that the gapless bound states are also protected by an orbital-exchange symmetry which is present for these junctions in the limit considered here, although the protection is not topological in origin. We find that breaking the orbital-exchange symmetry leads to a restoration of Andreev scattering which mixes electron- and hole-like states, which, in turn, leads to a gap. An extreme example of this symmetry breaking is provided by $s\tau_{3}-N-s$ junctions, which exhibit Andreev bound states. A second important feature is that the bound state spectra for the $s\tau_{3}-N-s\tau_{3}$ and $s\tau_{3}-N-s$ junctions are invariant to changes in the \emph{global} phase difference across the junction. 
This is also in striking contrast to the typical single-channel Josephson junction, 
where the Andreev bound state spectrum changes with phase, leading to a static Josephson effect. 
Our unusual findings are confirmed via microscopic calculations of the Josephson currents in the tunneling limit. We show in this limit that the contributions from the two sectors of $s\tau_{3}$ pairing, which have opposite intrinsic phases, cancel. 

We also consider more realistic junctions where the SC leads include intra- and inter-orbital hybridization. We focus on cases where the band splitting in the vicinity of the Fermi surface (FS) is much smaller than the pairing amplitudes. We show that intra-orbital hybridization terms break the orbital-exchange symmetry, leading to bound states more akin to the usual, gapped Andreev bound states. The same terms also make the bound state spectrum sensitive to a global phase difference across the junction. These findings are further confirmed by microscopic calculation of the Josephson current in the tunneling limit. As a counter-example to the gapped bound state spectrum,  we consider setups where the junction is along the $z$-direction, while the intra- and inter-orbital hybridization are predominantly in-plane. We show that gapless states can still be found for in-plane momenta along the diagonals, where the orbital-exchange symmetry is preserved. In all cases, we show that the analytical solutions are consistent with numerical results. Fig.~\ref{Fig:Drft_Schmts_v1} provides a summary of our main results.    

As discussed in Refs.~\cite{Nica_Yu, Nica_Si_npj_2021}, $s\tau_{3}$ pairing is expected to provide a superconducting state whose single-particle excitations are gapped everywhere on the FS.
 The simplest case to consider is well established for the alkaline Fe-selenides. In this case, the splitting of the relevant bands near the FS, which is essentially determined by intra- and inter-orbital hybridization terms, is small when compared to the pairing amplitudes. This is likely the case for a similar proposal in CeCu$_{2}$Si$_{2}$~\cite{Nica_Si_npj_2021}. In light of our preceding discussion, the band-splitting in alkaline Fe-selenides and CeCu$_{2}$Si$_{2}$ necessarily leads to gapped bound states and to a finite Josephson current for arbitrary global phase differences between the two leads. However, we expect that these effects are small. Therefore, $s\tau_{3}-N-s\tau_{3}$ and $s\tau_{3}-N-I$ junctions will typically exhibit bound states with gaps which are much smaller than the bulk gap, although exceptions with gapless states are also possible for arbitrary global phase differences. Likewise, $s\tau_{3}-N-s\tau_{3}$ and $s\tau_{3}-N-s$ junctions will exhibit static Josephson currents which are determined by band-splitting near the FS, and are therefore small when compared to similar junctions 
 with orbitally-trivial pairing states. While these predictions do not provide unequivocal probes for $s\tau_{3}$ pairing, their combined signatures can provide significant supporting evidence.   

The remainder of the article is divided into the following sections. In Sec.~\ref{Sec:Mdls}, we introduce the microscopic models for the junctions. In Sec.~\ref{Sec:st3}, we discuss both analytical and numerical solutions for $s\tau_{3}-N-s\tau_{3}$ junctions, and we briefly compare these to  junctions with orbitally-trivial pairing. $s\tau_{3}-N-I$ junctions have similar bound states, and are briefly discussed at the end of the section. Section~\ref{Sec:s_trvl} is devoted to $s\tau_{3}-N-s$ junctions. In the final Sec.~\ref{Sec:Dscs}, we summarize our findings and discuss possible experimental realizations of the junctions. The appendices contain detailed discussions of some of the points presented in the main text. In Appendix~\ref{App:st3_N_st3}, we present the analytical bound state solutions for $s\tau_{3}-N-s\tau_{3}$ junctions in the important limit of degenerate orbitals. Appendix~\ref{App:Unql} briefly touches on the effect of unequal coupling of the two orbitals to the metallic part, while Appendix~\ref{App:st0} discusses junctions with orbitally-trivial pairings. Appendices~\ref{App:I} and~\ref{App:st3_N_s} present the analytical solutions for $s\tau_{3}-N-I$ and $s\tau_{3}-N-s$ junctions, respectively, both in the limit of degenerate orbitals. The final Appendix~\ref{App:Jsph}, presents our results for the Josephson current for $s\tau_{3}-N-s\tau_{3}$ junctions, as well as for junctions with orbitally-trivial pairing.  

\section{Models}

\label{Sec:Mdls}

The main distinction of the setups considered here are the superconducting leads which are in a $s\tau_{3}$ pairing state, which was originally proposed for an effective $d_{xz}, d_{yz}$ two-orbital model for the alkaline Fe-selenides~\cite{Raghu,Si_Abrahams2008,Daghofer2010}. 
The \emph{spin-singlet} pairing state consists of a $s_{x^{2}y^{2}}(k_{x}, k_{y})$ form factor multiplied 
by a $\tau_{3}$ Pauli matrix in orbital space. Due to the non-trivial $\tau_{3}$ orbital structure, $s\tau_{3}$ pairing transforms as a $B_{1g}$ irreducible representation of the tetragonal $D_{4h}$ point-group, 
and thus it changes sign under a $\pi/2$ rotation about $z$. In the simplest case, with the pairing amplitude 
being larger than the band-splitting near the FS, 
which is due to intra- and inter-orbital hybridization terms in the normal-state Hamiltonian, the Bogoliubov-de Gennes (BdG) spectrum of this state is 
\emph{fully gapped}. In the more general cases, it is always gapped on the FS. In Ref.~\cite{Nica_Yu}, $s\tau_{3}$ pairing was stabilized in a realistic five-orbital model for the alkaline Fe-selenides. 

When considering Josephson junctions, the gapped nature of the $s\tau_{3}$ state implies that the relative phases of the pairing of the two $d_{xz}, d_{yz}$ orbital sectors are preserved by the junction. 
Furthermore, the common $s_{x^{2}y^{2}}$ form factor has nodes along the $k_{x}= \pm \pi/2$ and $k_{y}= \pm \pi/2$ axes. For FS's away from these axes, we can ignore the momentum-dependence of the pairing, which does not introduce any qualitatively new effects. 

\subsection{Junctions along $x$}

\label{Sec:Mdls_st3_x}

For all of the three types of junctions along $x$, we adopt the following model

\enni \begin{align}
H = & H_{\text{L}} + H_{\text{L-C}} 
+ 
H_{\text{C}} 
+ 
H_{\text{C-R}}
+ 
H_{\text{R}}.
\end{align}

\enni $H_{\text{L}}$, $H_{\text{R}}$, and $H_{\text{C}}$ are the bulk Hamiltonians for the left, right leads and the center metallic parts, respectively. $H_{\text{L-C}}$ and $H_{\text{C-R}}$ contain the terms at the lead-center part interfaces, and by convention, are defined on the last and first sites of the L and R leads, respectively.
The L lead which includes the pairing terms
extends along  $x[a] \le 
(L_{x} -l)/2$, where $L_{x}$ is the number of sites for
   the entire L-C-R system , while $l$ is the number of sites of the metallic C part, along the $x$-direction, respectively. The length is defined in units of the lattice spacing $a$. The most general form of the bulk L lead Hamiltonian for $s\tau_{3}$ pairing is 

\enni \begin{align}
H_{\text{L}} = & H_{\text{L, TB}} + H_{\text{L, Pair}}, ~\text{for}~
x[a]
< 
\left( 
\frac{L_{x}-l}{2} - 1
\right).
\end{align}

\begin{widetext}

\enni \begin{align}
H_{\text{L, TB}}
= & \sum_{\alpha \textbf{r}\sigma} 
\bigg[
 -\left( t_{x \alpha} c^{\dag}_{\mathbf{r}, \alpha \sigma} c_{\mathbf{r}+ \hat{\mathbf{x}}, \alpha \sigma} 
+ 
t_{y \alpha} c^{\dag}_{\mathbf{r}, \alpha \sigma} c_{\mathbf{r}+ \hat{\mathbf{y}}, \alpha \sigma} 
+\text{H.c.} \right) 
- \mu c^{\dag}_{\mathbf{r}, \alpha \sigma} c_{\mathbf{r}, \alpha \sigma}
\notag \\
+ & \sum_{\beta \neq \alpha}
t_{4} \left( 
c^{\dag}_{\mathbf{r}, \alpha \sigma} c_{\mathbf{r}+\hat{\mathbf{x}} + \hat{\mathbf{y}}, \beta \sigma} 
-
c^{\dag}_{\mathbf{r}, \alpha \sigma} c_{\mathbf{r}+\hat{\mathbf{x}} - \hat{\mathbf{y}}, \beta \sigma} + \text{H.c.} 
\right)
\bigg] 
\label{Eq:H_LTB}
\end{align}

\end{widetext}

\enni \begin{align} 
H_{\text{L, Pair}}=  \sum_{\alpha \mathbf{r}}  \Delta_{\alpha, \text{L}} 
\left( 
c^{\dag}_{\mathbf{r}, \alpha \uparrow} c^{\dag}_{\mathbf{r}, \alpha \downarrow}
- 
c^{\dag}_{\mathbf{r}, \alpha \downarrow} c^{\dag}_{\mathbf{r}, \alpha \uparrow}
\right)
 + \text{H.c.}
\end{align}

\enni The indices $\alpha \in \{1, 2 \}$ stand for the $d_{xz}$ and $d_{yz}$ orbitals, respectively, while the spin is represented by $\sigma \in \{\uparrow, \downarrow \}$. $t_{x\alpha}$ are the nearest-neighbor (NN) hopping coefficients for orbital $\alpha$, while $t_{4}$ is a NN orbital hybridization, with all coefficients taken to be real. These are based on a simplified two-orbital model introduced in Ref.~\cite{Raghu}. Note the absence of terms proportional to a $\tau_{2}$ matrix in the tight-binding part, since these break time-reversal symmetry. The pairing satisfies $ \Delta_{1}=- \Delta_{2}= \Delta$. As mentioned previously, we neglect the spatial dependence of the pairing which is not of immediate importance to the effects considered here.

For periodic boundary conditions (BC's), 
the BdG Hamiltonian for the L lead and for single spin sector reads

\enni \begin{align}
H_{\text{L}}(\mathbf{k}) =
\begin{pmatrix}
\xi_{0} + \xi_{3}  & \xi_{1} & \Delta & 0 
\\
\xi_{1} & \xi_{0} - \xi_{3}  & 0 & -\Delta 
\\
 \Delta^{*} & 0 &  -(\xi_{0} + \xi_{3} ) & -\xi_{1}
 \\
 0 & -\Delta^{*} & - \xi_{1} & -(\xi_{0} - \xi_{3} )
\end{pmatrix}
\label{Eq:BdG_st3_ld}
\end{align} 

\enni where 

\enni \begin{align}
\xi_{0}(\mathbf{k}) = & (t_{1} + t_{2})  \left[ 
\cos(k_{x}a) + \cos(k_{y}a) 
\right] - \mu 
\label{Eq:xi0}
\\
\xi_{3}(\mathbf{k}) = & (t_{1} - t_{2})  \left[ 
\cos(k_{x}a)- \cos(k_{y}a)
\right]
\label{Eq:xi3}
\\
\xi_{1}(\mathbf{k}) = & -4t_{4} \sin(k_{x}a) \sin(k_{y}a).
\label{Eq:xi1}
\end{align}

\enni $D_{4h}$ symmetry restricts 

\enni \begin{align}
t_{x1} = & t_{y2} =  t_{1} \label{Eq:t1}\\
t_{x2} = & t_{y1} =  t_{2} \label{Eq:t2}.
\end{align}

\enni Three terms $\xi_{0} \tau_{0}, \xi_{1} \tau_{1}$, and $\xi_{3} \tau_{3}$ determine the normal state, corresponding to a common dispersion, inter- and intra-orbital hybridization terms, respectively. Note that terms $\propto \cos(k_{x}a) \cos(k_{y}a)$, which preserve the lattice symmetry can also be added to the orbital-\emph{diagonal} $\xi_{0}$ terms~\cite{Raghu}. While these can lead to a change in shape of the FS, they preserve the orbital structure of $H_{\text{L}}$, i.e. do not induce any additional band splitting, which is essential to the results of this work. Therefore, we ignore these additional contributions. The important degenerate-orbital limit, which we mention in the following, occurs when both inter- and intra-orbital hybridization terms are set to zero with $\xi_{1} =\xi_{3} =0$. For junctions along $x$ which break the translation symmetry along this direction, this corresponds to $t_{1} - t_{2} = t_{4} =0.$

The positive eigenvalues are

\begin{align}
E_{1,2} = 
\sqrt{
\xi^{2}_{0} + \xi^{2}_{3}
+ \xi^{2}_{1} + |\Delta|^{2}
\pm 2 
\sqrt{\xi^{2}_{0} 
\left(\xi^{2}_{1} + \xi^{2}_{3}
\right)
+ \xi^{2}_{1} |\Delta|^{2}}
}.
\end{align}

\enni As mentioned previously, we consider cases where the band splittings near either of the two FS's are smaller than the pairing amplitude

\enni \begin{align}
\sqrt{\xi^{2}_{1}(\mathbf{k}) + \xi^{2}_{3}(\mathbf{k})} \big|_{\mathbf{k} \in \text{FS}} < |\Delta|^{2}. 
\end{align}

\enni which ensures that the leads are gapped in the bulk. 

The Hamiltonian at the L-C interface is 

\begin{widetext}

\enni \begin{align}
H_{\text{L-C}}= &  \sum_{\alpha y} 
\bigg\{ 
 \sum_{\sigma} 
\left[ 
 -\left( V_{\alpha} c^{\dag}_{\mathbf{r}, \alpha  \sigma} 
 c_{ \mathbf{r} + \hat{\mathbf{x}}, \sigma} 
+ t_{y \alpha} c^{\dag}_{\mathbf{r}, \alpha \sigma} c_{\mathbf{r} + \hat{\mathbf{y}}, \alpha \sigma}
+ \text{H.c.} \right) 
 - \mu c^{\dag}_{\mathbf{r}, \alpha \sigma} c_{\mathbf{r}, \alpha \sigma} 
 \right]
 + 
 \left[ \Delta_{\alpha} 
\left( 
c^{\dag}_{\mathbf{r}, \alpha \uparrow} c^{\dag}_{\mathbf{r}, \alpha \downarrow}
- 
c^{\dag}_{\mathbf{r}, \alpha \downarrow} c^{\dag}_{\mathbf{r}, \alpha \uparrow}
\right)
 + \text{H.c.}
 \right]
 \bigg\}
 \notag \\
 ,~& \text{for}~x
 [a]
 = 
 \frac{L_{x}-l}{2} 
 \label{Eq:HLC}
\end{align}

\end{widetext}

\enni 
As mentioned earlier, the $k$-dependence of the $s(k)$ factor in the $s\tau_{3}$ pairing is unimportant for our purpose, and this is 
reflected in the form of the
pairing term of  Eq.~\ref{Eq:HLC}.
Crucially, the terms proportional to $V_{\alpha}$ denote the hybridization of the two orbitals in the L lead to \emph{a single orbital} in the C part along $x$. Since the point-group symmetry is necessarily broken at the lead-center interfaces, this symmetry does not \emph{a priori} restrict the values of the $V_{\alpha}$'s. In the following, we allow these couplings to take arbitrary values and we comment on the effects of the $d_{xz}, d_{yz}$ nature of the orbitals in the leads in Sec.~\ref{Sec:Dscs}. The remaining terms are identical to those of $H_{\text{L}}$ while the summation is over the $y$ coordinate along the junction. We typically choose $V_{1}=V_{2}=t_{1}=1$, unless stated otherwise. 
As already mentioned, we neglect the spatial dependence of the pairing.

The Hamiltonian for the bulk of the central (C) part reads

\begin{widetext}

\enni \begin{align}
H_{\text{C}}= &  \sum_{\alpha \mathbf{r} \sigma} 
\left[
-t_{1} \left(  c^{\dag}_{\mathbf{r}, \sigma} c_{\mathbf{r} + \hat{\mathbf{x}}, \sigma} 
+ 
c^{\dag}_{\mathbf{r}, \sigma} c_{\mathbf{r} + \hat{\mathbf{y}}, \sigma}
+\text{H.c.} \right)
 - \mu  c^{\dag}_{\mathbf{r}, \sigma} c_{\mathbf{r}, \sigma}
 \right],~\text{for} 
~\frac{L_{x}-l}{2} +1  \le 
x[a]
 \le \frac{L_{x}+l+2}{2}  - 1.
\end{align}

\end{widetext}

\enni As mentioned previously, the C part involves a \emph{single channel} without any pairing. We consider for simplicity nearest-neighbor hopping in this sector, taken to be equal to $t_{1}$. 

In the case of $s\tau_{3}-N-s\tau_{3}$ pairing we choose the Hamiltonians at the C-R junction $H_{\text{C-R}}$ for 
$x[a] = (L_{x}+l+2)/2$ and for the bulk of the R lead
$H_{\text{R}}$ for 
$x[a] \ge (L_{x}+l+2)/2 + 1 $
to have identical form and coefficients to $H_{\text{L-C}}$ and $H_{\text{L}}$, respectively, with the exception of a possible \emph{global non-zero phase} $\phi$ for the pairing terms: 

\enni \begin{align}
\Delta_{\text{L}} = & \Delta 
\\
\Delta_{\text{R}} = & \Delta e^{i \phi}.
\end{align}

For $s\tau_{3}-N-s$, $H_{\text{C-R}}$ and $H_{\text{R}}$ reflect the presence of a single channel in the R lead as

\enni \begin{align}
H_{\text{C-R}}= &  \sum_{\alpha y} 
\bigg\{
 \sum_{\sigma} 
 \bigg[ 
 -\left( V c^{\dag}_{\mathbf{r}, \sigma} 
 c_{ \mathbf{r} - \hat{\mathbf{x}}, \sigma} 
+ t_{y} c^{\dag}_{\mathbf{r}, \sigma} c_{\mathbf{r} + \hat{\mathbf{y}}, \sigma}
+ \text{H.c.} \right) 
\notag \\
 - & \mu c^{\dag}_{\mathbf{r}, \alpha \sigma} c_{\mathbf{r}, \alpha \sigma} 
 \bigg]
 + 
 \left[ \Delta e^{i \phi}
\left( 
c^{\dag}_{\mathbf{r}, \uparrow} c^{\dag}_{\mathbf{r},  \downarrow}
- 
c^{\dag}_{\mathbf{r}, \downarrow} c^{\dag}_{\mathbf{r},  \uparrow}
\right)
 + \text{H.c.}
 \right]
 \bigg\} 
 \notag \\
  &, ~\text{for}~x[a]
  = \frac{L_{J}+l+2}{2} 
\end{align}

\enni \begin{align}
H_{\text{R, TB}}= & - \sum_{\textbf{r} \sigma} 
\bigg[
-t  \left( c^{\dag}_{\mathbf{r}, \sigma} c_{\mathbf{r}- \hat{\mathbf{x}}, \sigma} 
+ 
c^{\dag}_{\mathbf{r}, \alpha \sigma} c_{\mathbf{r}+ \hat{\mathbf{y}}, \sigma} 
+\text{H.c.} \right) 
\notag \\
- & \mu c^{\dag}_{\mathbf{r}, \sigma} c_{\mathbf{r}, \sigma}
\bigg]
\end{align}

\enni \begin{align} 
H_{\text{R, Pair}}=  \sum_{\mathbf{r}}  \Delta  e^{i \phi}
\left( c^{\dag}_{\mathbf{r},  \uparrow} c^{\dag}_{\mathbf{r}, \downarrow} 
-
c^{\dag}_{\mathbf{r},  \downarrow} c^{\dag}_{\mathbf{r}, \uparrow} 
\right)
 + \text{H.c.}.
\end{align}

\enni For simplicity, we always consider $V=t=1$. 

\subsection{Junctions along $z$}

\label{Sec:Mdls_st3_z}

For $s\tau_{3}-N-s\tau_{3}$ junctions along $z$ we define the complete Hamiltonian via 

\enni \begin{align}
\tilde{H} = & \tilde{H}_{\text{L}} + \tilde{H}_{\text{L-C}} 
+ 
\tilde{H}_{\text{C}} 
+ 
\tilde{H}_{\text{C-R}}
+ 
\tilde{H}_{\text{R}}.
\end{align}

\enni In addition to the terms already present in the pure two-dimensional case, we consider additional NN hopping along $z$ for the L/R leads as 

\enni \begin{align}
\tilde{H}_{\text{L, TB}}= & H_{\text{L, TB}} - \sum_{\alpha \textbf{r} \sigma} 
 t_{z} \left(  c^{\dag}_{\mathbf{r}, \alpha \sigma} c_{\mathbf{r}+ \hat{\mathbf{z}}, \alpha \sigma} 
+\text{H.c.} \right) ,
\end{align}

\enni where $H_{\text{L, TB}}$ was defined in Eq.~\ref{Eq:H_LTB}, and the sum over lattice sites is now along all of the three axes. We take the lengths of the L, C, and R parts along $z$ to be the same as for the junctions along $x$. For simplicity, we consider only orbital-diagonal, NN hopping along the $z$-direction. Additional terms are of course possible, but they do not qualitatively change our conclusions. For clarity, we ignore the momentum dependence of the two-dimensional pairing, as for junctions along $x$. Similar expressions hold for the bulk of the R lead, with
the exception of a global phase difference $\phi$.  Likewise, the C part amounts to 

\enni \begin{align}
\tilde{H}_{\text{C}}= & H_{\text{C}} - \sum_{\alpha \textbf{r}}  
 t_{z} \left(  c^{\dag}_{\mathbf{r}, \sigma} c_{\mathbf{r}+ \hat{\mathbf{z}}, \sigma} 
+\text{H.c.} \right).
\end{align}

\enni We use a NN hopping in the C part which is equal in amplitude to that of the L/R leads $t_{z}=0.2t_{1}$, unless stated otherwise. The reduced value of $t_{z}$ reflects the tetragonal anisotropy of our target systems. 

Finally, the Hamiltonians for the L-C and C-R interfaces are obtained via a straightforward generalization to hopping along $z$

\begin{widetext}

\enni \begin{align}
H_{\text{L-C}}= &
- \sum_{\alpha, xy}
\bigg\{
\sum_{\sigma}
\bigg[ 
\left(- V_{\alpha z} c^{\dag}_{\mathbf{r}, \alpha \sigma} 
 c_{\mathbf{r} + \hat{\mathbf{z}}, \sigma} 
-  t_{x \alpha} c^{\dag}_{\mathbf{r}, \alpha \sigma} c_{\mathbf{r}+ \hat{\mathbf{x}}, \alpha \sigma} 
- 
t_{y \alpha} c^{\dag}_{\mathbf{r}, \alpha \sigma} c_{\mathbf{r}+ \hat{\mathbf{y}}, \alpha \sigma} 
+\text{H.c.} \right) 
- \mu c^{\dag}_{\mathbf{r}, \alpha \sigma} c_{\mathbf{r}, \alpha \sigma}
\notag \\
+ & 
\sum_{\beta \neq \alpha} 
t_{4} \left( 
c^{\dag}_{\mathbf{r}, \alpha \sigma} c_{\mathbf{r}+\hat{\mathbf{x}} + \hat{\mathbf{y}}, \beta \sigma} 
-
c^{\dag}_{\mathbf{r}, \alpha \sigma} c_{\mathbf{r}+\hat{\mathbf{x}} - \hat{\mathbf{y}}, \beta \sigma} + \text{H.c.} 
\right)
\bigg]  
+ \left[ 
\Delta_{\alpha} 
\left( 
c^{\dag}_{\mathbf{r}, \alpha \uparrow} c^{\dag}_{\mathbf{r}, \alpha \downarrow}
- 
c^{\dag}_{\mathbf{r}, \alpha \downarrow} c^{\dag}_{\mathbf{r}, \alpha \uparrow}
\right)
 + \text{H.c.} 
 \right]
 \Bigg\}
\end{align}

\end{widetext}

\enni $\tilde{H}_{\text{C-R}}$ can be obtained via a similar generalization of $H_{\text{C-R}}$. 

Unless stated otherwise, we choose units where the nearest-neighbor hopping and hybridization at the interfaces are $t_{1}=V_{1}=1$ and $\Delta=0.4$.

\subsection{Orbital-exchange symmetry}

\label{Sec:Orbl_exch}

We consider the $s\tau_{3}-N-s\tau_{3}$ and $s\tau_{3}-N-I$ setups in the important limit where the two orbitals have identical tight-binding coefficients, zero intra-orbital hybridization ($\xi_{3}=0$) and inter-orbital hybridization ($\xi_{1}=0$), and couple identically to the single orbital of the C part. This limit corresponds to 

\enni \begin{align}
t_{1} = & t_{2} \label{Eq:Dgnr_1}\\
t_{4} = & 0 \\
V_{1} = & V_{2} \label{Eq:Dgnr_3}.
\end{align} 

\enni  The two orbitals in the L (or R  if appropriate) lead correspond to degenerate bands in the normal state. We define a \emph{local} transformation which acts on the L (and R)  spinors $\Psi^{T}= (c_{\mathbf{r}, 1 \up}, c_{\mathbf{r}, 2 \up}, c^{\dag}_{\mathbf{r}, 1 \dn}, c^{\dag}_{\mathbf{r}, 2 \dn})$ as

\enni \begin{align}
\Psi \rightarrow 
\hat{R}
\Psi 
\end{align}

\enni where 
 
\enni \begin{align}
\hat{R} =
\begin{pmatrix}
0 & i & 0 & 0 \\
i & 0 & 0 & 0 \\
0 & 0 & 0 & -i \\
0 & 0 & -i & 0
\end{pmatrix}.
\end{align}

\enni This corresponds to an exchange of the two orbitals followed by a gauge transformation which changes the sign of the pairing $\Delta$. Similarly, in the C part the transformation acting on the single-orbital spinor $\psi^{T}= (c_{\mathbf{r}, \up}, c^{\dag}_{\mathbf{r} , \dn})$

\enni \begin{align}
\psi \rightarrow \hat{S} \psi
\end{align}

\enni where 

\enni \begin{align}
\hat{S} = 
\begin{pmatrix}
i & 0  \\
0 & -i \\
\end{pmatrix}.
\end{align}

\enni This operation, corresponding to a gauge transformation for the C part, is required to compensate for the factors of $i$ in $H_{\text{L-C}}$ and $H_{\text{C-R}}$ if appropriate. Both $\hat{R}$ and $\hat{S}$ are anti-hermitian operators which obey

\enni \begin{align}
\hat{R}^{\dag} = & - \hat{R} \\
\hat{R}^{2} = & - \hat{1}
\end{align}

\enni and similarly for $\hat{S}$. 

It is straightforward to define an orbital-exchange operator for the entire lattice models, corresponding to each of the junctions. For $s\tau_{3}-N-s\tau_{3}$ junctions it reads

\enni \begin{align}
\hat{Q} =
\begin{cases}
\hat{R},~\forall~x \in \text{L, R} \\
\hat{S},~\forall~x \in \text{C}
\end{cases}
\end{align}

\enni The $s\tau_{3}-N-s\tau_{3}$ and $s\tau_{3}-N-I$ junction Hamiltonians in the degenerate-orbital limit considered here are invariant under $\hat{Q}$. This orbital exchange symmetry plays an important role in classifying the electron-like and hole-like solutions encountered in these cases. 

\section{$s\tau_{3}-N-s\tau_{3}$ and $s\tau_{3}-N-I$ junctions}

\label{Sec:st3}

We treat $s\tau_{3}-N-s\tau_{3}$ junctions in some detail in the following and only briefly discuss $s\tau_{3}-N-I$ junctions at the end of the section and in Appendix~\ref{App:st0}, since the latter exhibit similar bound states.

\subsection{Degenerate orbitals}

\label{Sec:St3_dgnr}

We consider a $s\tau_{3}-N-s\tau_{3}$ junction as defined previously, in the limit of degenerate orbitals 
( Eqs.~\ref{Eq:Dgnr_1}-\ref{Eq:Dgnr_3} ).
 We find the bound states in the continuum limit by linearizing the BdG equations in the vicinity of points $(\alpha K_{Fx}, K_{Fy})$ on the FS~\cite{Sauls_2018}, where $\alpha=\pm 1$. The detailed solution is presented in Appendix~\ref{App:st3_N_st3}. Here, we summarize some of the most important results. In contrast to the typical single-channel junction, the presence of two orbitals in either leads, which couple to a single orbital in the C part, imposes additional BC's. Indeed, in the limit considered here, only the linear combination $c_{1}+ c_{2}$ of the two orbitals couples to the C part. The corresponding BdG coefficients must vary continuously at the L-C or C-R interfaces, much like in a single-channel junction. By contrast, the anti-symmetric combination does not couple to the C part and thus the corresponding BdG coefficients obey open BC's at the interface. As shown in Appendix~\ref{App:st3_N_st3}, the two channels,  corresponding to symmetric and anti-symmetric linear combinations, do not decouple in the bulk of the leads, due to the non-trivial matrix structure of $s\tau_{3}$ pairing. Consequently, the continuity and open BC's cannot be satisfied simultaneously for solutions which involve both electron- and hole-like solutions in the C part. The solutions are either electron- or hole-like in the C part, in contrast to the typical Andreev bound states which mix the two.

The electron-like bound states with eigenvalues 

\enni \begin{align}
\frac{\epsilon}{|\Delta|} = & \pm 
\cos\left( \frac{\epsilon l}{2 v_{Fx}} + \frac{K_{Fx}l}{2} + \frac{(a-m) \pi}{2} \right). 
\label{Eq:Elct_enrg}
\end{align}

\enni have the form

\enni \begin{align}
\Psi^{L/R}_{K_{Fy};\text{Electron}} = & \left(
u^{L/R}_{1e}, u^{L/R}_{2e}, v^{L/R}_{1e}, v^{L/R}_{2e} \right)^{T} 
\label{Eq:Spnr_elct}
\end{align}

\enni \begin{align}
u^{L/R}_{1e} = & \frac{1}{\sqrt{2}}|A| |\Delta| e^{i \theta^{0}_{A}} 
\sin \left[ K_{Fx} x  \mp \frac{\epsilon l}{2 v_{Fx}} + \frac{(a+m)\pi} {2} \right]
\\
u^{L/R}_{2e} = & u^{L/R}_{1} 
\\
v^{L/R}_{1e} = & \frac{1}{\sqrt{2}} |A| |\Delta| e^{i \theta^{0}_{A}} \sin \left[ K_{Fx} \left( x \pm \frac{l}{2} \right)+ \gamma_{L/R} \pi \right] 
\\
v^{L/R}_{2e} = & - v^{L/R}_{1} \\
\gamma_{L}= & a \\
\gamma_{R} = & m
\end{align}

\enni in the L/R leads and 

\enni \begin{align}
\Psi^{C}_{K_{Fy};\text{Electron}} = & \left(
u^{C}_{e}, v^{C}_{e} \right)^{T} 
\end{align}

\enni \begin{align}
u^{C}_{e} = & 2|A| |\Delta| e^{i \theta^{0}_{A}} 
\sin \left[ K_{Fx} x + \frac{\epsilon x}{v_{Fx}} + \frac{(a+m)\pi}{2}  \right] 
\\
v^{C}_{e} = & 0. 
\end{align}

\enni in the C part. $|A|$ is a normalization constant,  $\theta^{0}_{A}$ is an arbitrary phase, and $a, m$ are arbitrary integers. Crucially, the C $v$ BdG coefficient is identically zero. 

Similarly, the hole-like solutions with eigenvalues

\enni \begin{align}
\frac{\epsilon}{|\Delta|} 
= & \pm 
\cos\left( \frac{\epsilon l}{2 v_{Fx}} - \frac{K_{Fx}l}{2} + \frac{(n-b) \pi}{2} \right). 
\label{Eq:Hl_enrg}
\end{align}

\enni are of the form

\enni \begin{align}
\Psi^{L/R}_{K_{Fy};\text{Hole}} = & \left(
u^{L/R}_{1h}, u^{L/R}_{2h}, v^{L/R}_{1h}, v^{L/R}_{2h} \right)^{T} 
\label{Eq:Spnr_hl}
\end{align}

\enni \begin{align}
u^{L/R}_{1h} = & \frac{1}{\sqrt{2}} |B| e^{i \theta^{B}_{0}}  
\sin \left[ K_{Fx} \left( x \pm \frac{l}{2} \right) + \gamma_{L/R} \pi \right]
\\
u^{L/R}_{2h} = & -u^{L}_{1h} 
\\
v^{L/R}_{1h} = & \frac{1}{\sqrt{2}} |B| e^{i \theta^{B}_{0}}  \sin \left[ K_{Fx}x \pm \frac{\epsilon l}{2v_{Fx}} + 
 \frac{\pi}{2}+\frac{(n+b)\pi}{2} \right] 
\\
v^{L/R}_{2h} = & v^{L}_{1h} \\
\gamma_{L}= & b \\
\gamma_{R} = & n
\end{align}

\enni for the L/R leads and 

\enni \begin{align}
\Psi^{C}_{K_{Fy};\text{Hole}} = & \left(
u^{C}_{h}, v^{C}_{h} \right)^{T} 
\end{align}

\enni \begin{align}
u^{C}_{h} =& 0 \\ 
v^{C}_{h} = & \sin \left[ K_{Fx} x - \frac{\epsilon x}{v_{Fx}} +\frac{(n+b)\pi}{2} \right]
\end{align}

\enni in the C part. As before, $|B|$ is a normalization constant, $\theta^{0}_{B}$ is an arbitrary phase, while $b,n$ are integers. In contrast to the electron-like solutions, $u^{C}=0$. 

These detailed analytical solutions present a number of important features. Firstly, the solutions are either electron- or hole-like since the corresponding BdG coefficient $v$ or $u$ vanishes in the C part. This is in clear contrast to a single-channel junction, where electron-like and hole-like solutions are mixed via Andreev scattering. The presence of both open and continuity BC's at L-C and C-R interfaces, as well as the non-trivial structure of the pairing, prevents the existence of solutions which mix electron and hole states. 

Secondly, electron- and hole-like solutions become gapless when 

\enni \begin{align}
K_{Fx}l = & (m-a) \pi \label{Eq:Gpls_1} \\
K_{Fx}l = & (n-b) \pi, \label{Eq:Gpls_2}
\end{align}

\enni respectively, as indicated by Eqs.~\ref{Eq:Elct_enrg} and~\ref{Eq:Hl_enrg}. This happens when $K_{Fx}l$ is either 0 or a multiple of $\pi$. Recall that the solutions are labeled by the momenta $(K_{Fx}, K_{Fy})$ which vary continuously along the FS. Consequently, the gapless conditions can be realized in multiple instances along the FS if the latter and/or junction length $l$ are sufficiently large. Linearizing Eqs.~\ref{Eq:Elct_enrg} and~\ref{Eq:Hl_enrg} about these points indicates that the electron- and hole-like states are counter-propagating, as a function of the conserved momentum $K_{Fy}$. 

Thirdly, the gaplessness of the electron- and hole-like states is protected by the orbital-exchange symmetry defined in Sec.~\ref{Sec:Orbl_exch}. Indeed, it is straightforward to check that

\enni \begin{align}
\hat{R}\Psi^{L/R}_{K_{Fy};\text{Electron}} = & i \Psi^{L/R}_{K_{Fy};\text{Electron}} 
\\
\hat{R}\Psi^{L/R}_{K_{Fy};\text{Hole}} = & i \Psi^{L/R}_{K_{Fy};\text{Hole}}, 
\end{align}

\enni together with

\enni \begin{align}
\Psi^{L/R, \dag}_{K_{Fy};\text{Electron}} \hat{R} = & i \Psi^{L/R, \dag}_{K_{Fy};\text{Electron}}.
\end{align}

\enni The last relation follows from the anti-hermitian nature of $\hat{R}$. Similar relations hold for $\hat{S}$ acting on the C spinors. Together, these imply that any operator $\hat{O}$ added to the Hamiltonian of the junction, which commutes with $\hat{R}$ and $\hat{S}$, will not mix electron- and hole-like states via 

\enni \begin{align}
\braket{ \Psi^{\dag}_{\text{Electron}}| \hat{O} \hat{R} |\Psi_{\text{Hole}} } = \braket{ \Psi^{\dag}_{\text{Electron}}| \hat{R} \hat{O} | \Psi_{\text{Hole}} }
\end{align}

\enni which implies that 

\enni \begin{align}
(-i) \braket{ \Psi^{\dag}_{\text{Electron}} |\hat{O} | \Psi_{\text{Hole}}}
 = & (+i) 
 \braket{ \Psi^{\dag}_{\text{Electron}}| \hat{O} | \Psi_{\text{Hole}} }
\end{align}

\and and subsequently that 

\enni \begin{align}
\braket{ \Psi^{\dag}_{\text{Electron}} |\hat{O} | \Psi_{\text{Hole}}} =0.
\end{align}

\enni Note that $\hat{O}$ is not necessarily local at the lattice level. A similar reasoning can be applied to $\hat{S}$. Therefore, the orbital-exchange symmetry protects the gapless electron- and hole-like states, in analogy to the time-reversal operator for spin-polarized edge states in quantum spin Hall systems~\cite{Kane_Mele}, although we stress that the gapless states are not due to any topological property of the bulk in our cases. In practice, the orbital-exchange symmetry is preserved when no intra-orbital hybridization terms (such as $\xi_{3}$ defined in Eq.~\ref{Eq:BdG_st3_ld}) are present, although inter-orbital hybridization terms (such as $\xi_{1}$ in Eq.~\ref{Eq:BdG_st3_ld}) preserve the orbital-exchange symmetry.

Fourthly, the bound state spectrum is independent of the global relative phase $\phi$ between L and R leads, as indicated by Eqs.~\ref{Eq:Elct_enrg} and~\ref{Eq:Hl_enrg}. This is in clear contrast to the single-channel junctions, where the change in the Andreev bound state spectrum with relative phase is proportional to the static Josephson current~\cite{Sauls_2018}. In the $s\tau_{3}-N-s\tau_{3}$ case, a bound state spectrum which is insensitive to the relative phase indicates the absence of a static Josephson current. This is confirmed by microscopic calculations of the Josephson current in the tunneling approximation, discussed in detail in Appendix~\ref{App:Jsph}, in the degenerate orbital limit considered here. As shown there, the Josephson currents due to each orbital sector cancel due essentially to the $\pi$ phase difference between the two components of $s\tau_{3}$ pairing. 

Next, we proceed to confirm the analytical solutions via numerical solutions of the lattice model introduced in Sec.~\ref{Sec:Mdls}, in the degenerate-orbital limit. The results were obtained for a lattice of 100 sites along $x$ with a C part of 5 sites extending along $49 \le  x \le 53$, in units of the lattice constant $a$. 

\begin{figure}[t!]
\includegraphics[width=1.0\columnwidth]{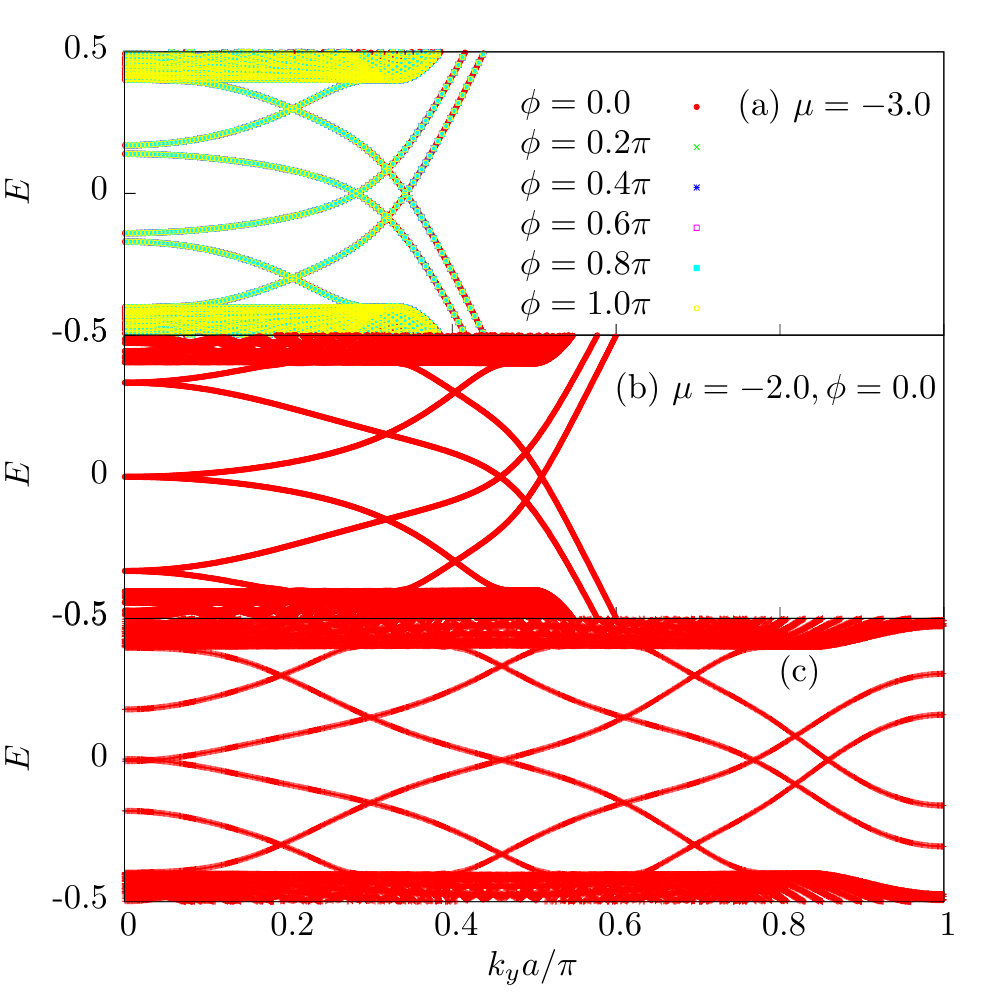}
\caption{Bound state spectrum of a $s\tau_{3}-N-s\tau_{3}$ junction along $x$, as defined in Sec.~\ref{Sec:Mdls_st3_x}, with degenerate orbitals in the L and R leads which also couple identically to a single orbital in the N part. (a) Eigenvalues for a fixed chemical potential $\mu=-3.0$ and varying global relative phase $\phi$ between the R and L superconducting leads, as functions of the conserved momentum $k_{y}$. Note that the spectrum is insensitive to $\phi$, in accordance to the analytical solution in Eqs.~\ref{Eq:Elct_enrg} and~\ref{Eq:Hl_enrg}. Also note the presence of gapless states near $k_{y}a \approx 0.3 \pi$.~(b) Same as (a) for fixed $\mu=-2.0$ and $\phi=0.0$. Although not shown, the spectrum is also insensitive to $\phi$. There is an additional crossing of gapless states near $k_{y}a=0.0$. This is consistent with the condition for gapless states obtained from the analytical solutions in Eqs.~\ref{Eq:Gpls_1}-\ref{Eq:Gpls_2}. With decreasing $\mu$ the FS increases, allowing $K_{Fx}l$ to reach values which are multiples of $\pi$ near $k_{y}=0$. (c) Same as (b) for $\mu=-0.25$ and $\phi=0$. Note that the number of gapless crossings increases with the size of the FS, as predicted by the analytical solution.}
\label{Fig:Drft_egnv_mu_vr}
\end{figure}

In Fig.~\ref{Fig:Drft_egnv_mu_vr}, we show the bound state spectrum as a function of the conserved momentum $k_{y}$, for varying chemical potential $\mu$ and global relative phase difference between the two superconducting leads $\phi$. For $\mu=-3.0$ in panel (a), we illustrate that the spectrum is independent of $\phi$, in accordance with the eigenvalues in Eqs.~\ref{Eq:Elct_enrg} and~\ref{Eq:Hl_enrg}. This implies the absence of static Josephson current, as confirmed in Appendix~\ref{App:Jsph} by a microscopic calculation of the Josephson current in the tunneling limit. Also note that the bound states become gapless and cross twice near $k_{y}a \approx 0.3 \pi$. This is consistent with the condition for gapless states as determined by the analytical solution (Eqs.~\ref{Eq:Gpls_1},~\ref{Eq:Gpls_2}). Indeed, for increasing chemical potential $\mu=-2.0$, as shown in panel (b), an additional crossing occurs near $k_{y}=0$. This is due to an $K_{Fx}$ (near $K_{Fy}=0$) which increases by $\pi$ as the FS expands. The crossings in (a) likewise shift to higher $k_{y}$. A similar picture presents itself with increasing chemical potential $\mu=-0.25$ in panel (c). Together, these results are consistent with the analytical solution.   

\begin{figure}[t!]
\includegraphics[width=1.0\columnwidth]
{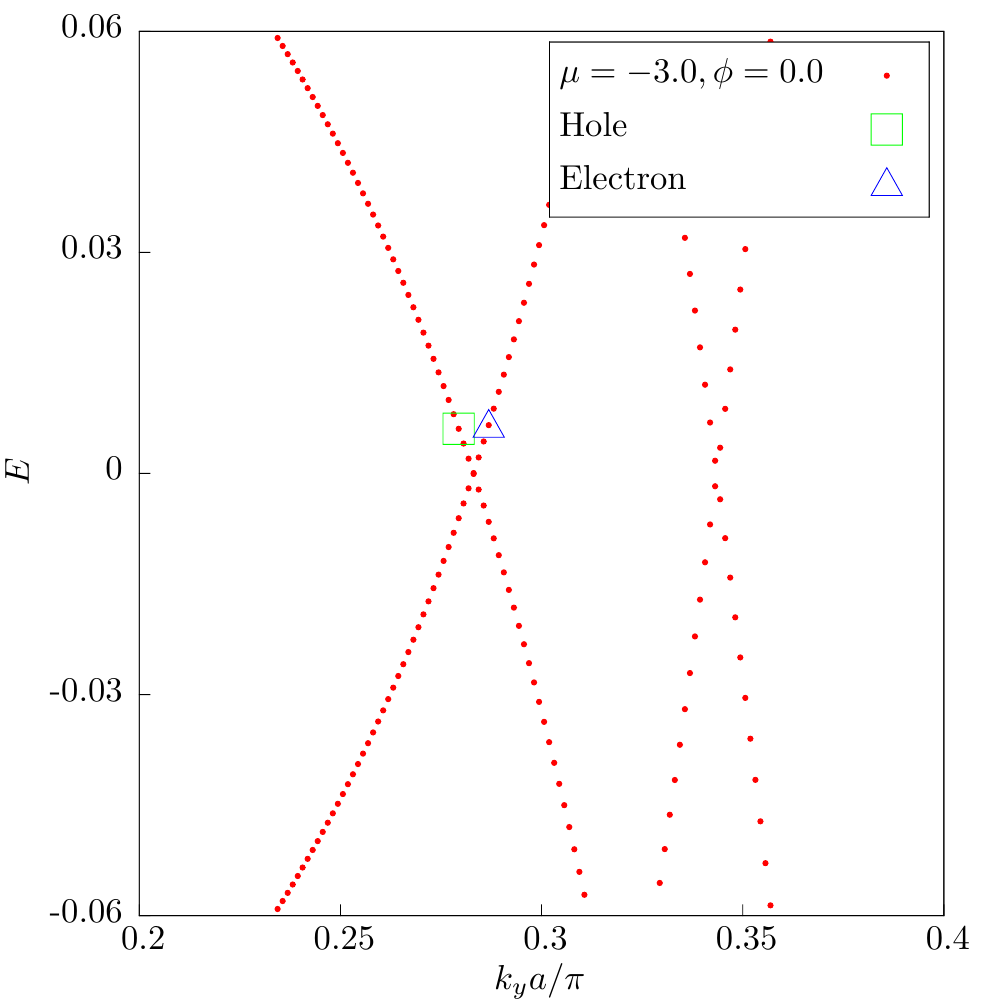}
\caption{Close-up of panel (a) of Fig.~\ref{Fig:Drft_egnv_mu_vr}~(a). The square and triangle identify the hole- and electron-like states illustrated in Fig.~\ref{Fig:Drft_ampl_cmpr_hl} and~\ref{Fig:Drft_ampl_cmpr_elct}, respectively.}
\label{Fig:Drft_Pstn_indc}
\end{figure}

\begin{figure}[t!]
\includegraphics[width=1.0\columnwidth]
{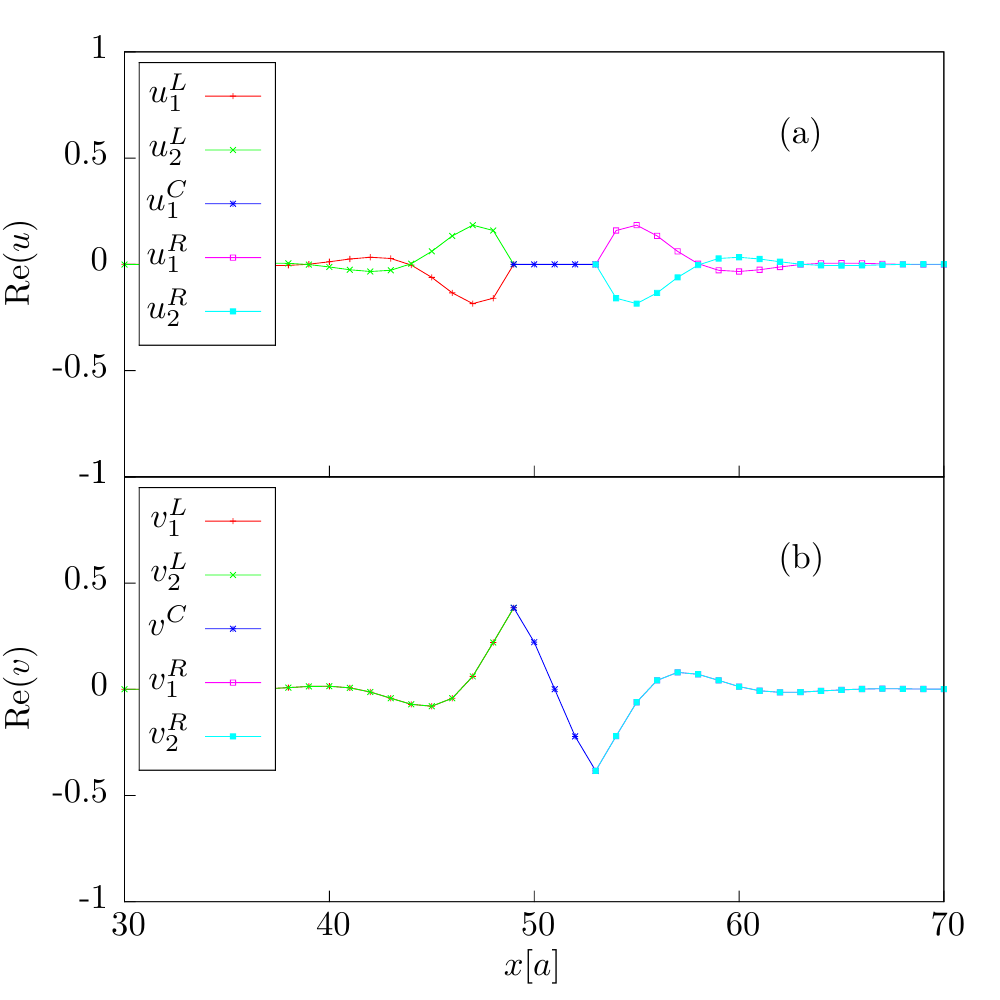}
\caption{Real parts of the BdG coefficients for a hole-like solution as a function of $x$ across the junction, for the momentum indicated by the square in Fig.~\ref{Fig:Drft_Pstn_indc}. These are obtained from the numerical calculations for $\mu=-3.0, \phi=0$. The imaginary parts vanish identically and are not shown. (a) The $u$ coefficients are consistent with Eq.~\ref{Eq:Spnr_hl}, as $u^{C}=0$ (blue squares) in the C part extending from $49  \le x[a]
 \le 53 $, as expected for a pure hole-like state. Similarly, $u^{L/R}_{1}$ and $u^{L/R}_{2}$ vanish at the L-C and C-R interfaces, respectively, in accordance to the analytical calculations. The opposite signs of these coefficients within the L and R leads and across the junction are also consistent with the analytical solution with $n-b$ an odd number. (b) The $v$ coefficients are finite throughout, including the C part, and are continuous across the junction. These  are also consistent with the analytical solution in Eq.~\ref{Eq:Spnr_hl}.}
\label{Fig:Drft_ampl_cmpr_hl}
\end{figure}

We also consider the nature of the bound states. In Fig.~\ref{Fig:Drft_Pstn_indc}, we show a close-up of panel (a) of Fig.~\ref{Fig:Drft_egnv_mu_vr} for $\phi=0$. One pair of hole- and electron-like states in the vicinity of a crossing are marked by a square and triangle, respectively. To elucidate the nature of these states, in Fig.~\ref{Fig:Drft_ampl_cmpr_hl} we illustrate the real parts of the BdG coefficients of the hole-like state marked in Fig.~\ref{Fig:Drft_Pstn_indc} by a square. Note that the imaginary pars are trivially zero and are not shown. The $u$ BdG coefficients shown in panel (a) are consistent with the analytical solution in Eq.~\ref{Eq:Spnr_hl}. Indeed, $u^{C}$ vanishes in the C part for $49 \le x[a] \le 53$, as expected for a hole-like solution. Furthermore, $u^{L/R}_{1}$ and $u^{L/R}_{2}$ both vanish at the L-C and C-R interfaces, respectively, and their signs are consistent with Eq.~\ref{Eq:Spnr_hl} for $n-b$ an odd multiple of $\pi$. In panel (b), the $v^{C}$ is finite and continuous across the junction. We conclude that the numerical solution is consistent with the hole-like solution indicated by the analytical results. 

\begin{figure}[t!]
\includegraphics[width=1.0\columnwidth]
{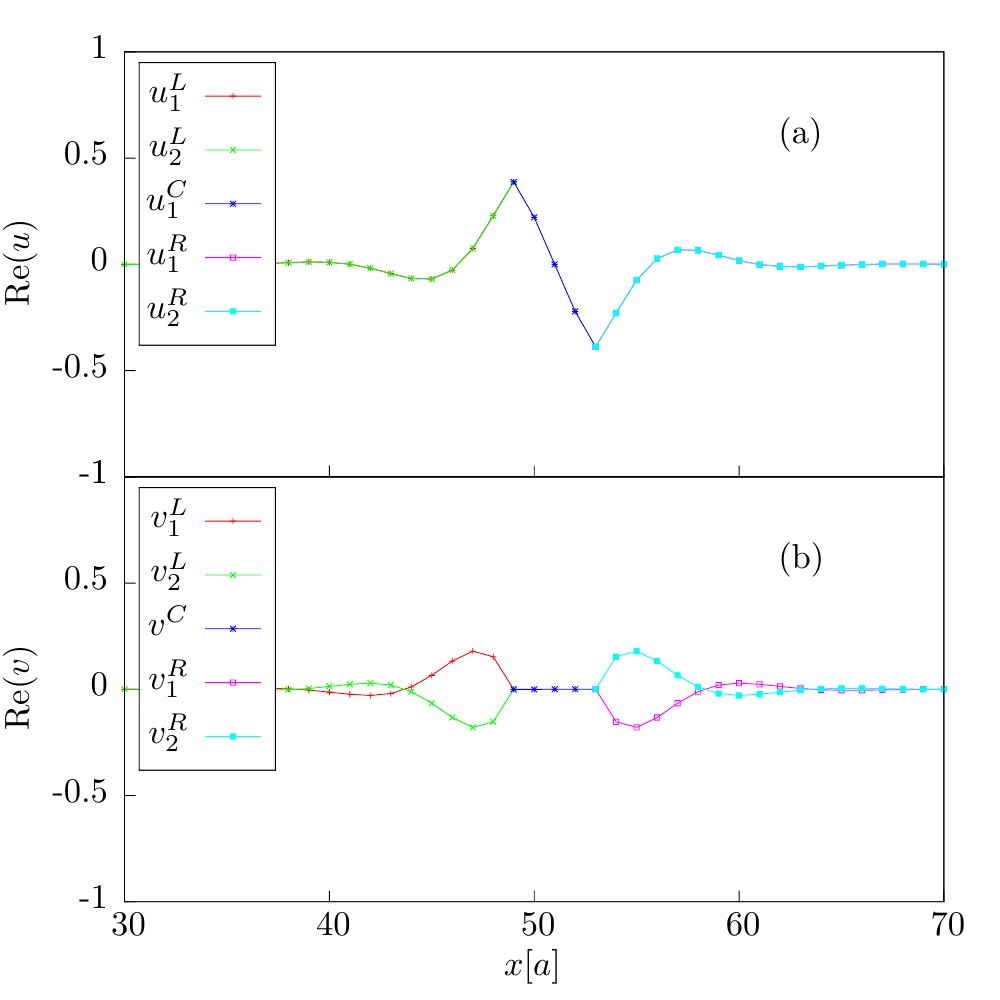}
\caption{Real part of the BdG coefficients for an electron-like solution as a function of $x$ across the junction, for the momentum indicated by the triangle in Fig.~\ref{Fig:Drft_Pstn_indc}. As in Fig.~\ref{Fig:Drft_ampl_cmpr_hl}, these are obtained from the numerical calculations for $\mu=-3.0, \phi=0$, and the imaginary parts vanish identically. (a) $u^{C}$  is finite, as expected for a pure electron-like state, and in agreement with the analytical solution in Eq.~\ref{Eq:Spnr_elct}. The remaining $u$ coefficients are also consistent with the latter. (b) $v^{C}=0$ again indicates a pure electron-like state. $v^{L/R}_{1}$ and $v^{L/R}_{2}$ vanish at the L-C and C-R interfaces, and their signs are consistent with $m-a$ and odd, as indicated by Eq.~\ref{Eq:Spnr_elct}.}
\label{Fig:Drft_ampl_cmpr_elct}
\end{figure}

A similar picture emerges for the electron-like state marked with a triangle in Fig.~\ref{Fig:Drft_Pstn_indc}. In Fig.~\ref{Fig:Drft_ampl_cmpr_elct}, we show the real parts of the BdG coefficients of this state across the junction, and compare these with the analytical solution in Eq.~\ref{Eq:Spnr_elct}. $u^{C}$ is now finite, while $v^{C}$ vanishes in the C region, as expected. Furthermore, the signs of $u^{L/R}_{1}$ and $u^{L/R}_{2}$ are consistent with a solution with $m-a$ an odd multiple of $\pi$. Together with the BdG coefficient for the hole-like state, these numerical solutions confirm the analytical results. 

We briefly discuss the case for a $s\tau_{3}-N-s\tau_{3}$ junction along $z$ for the model introduced in Sec.~\ref{Sec:Mdls_st3_z}. The analytical solution is similar to the junction along $x$ with the exception that both $k_{x}, k_{y}$ are conserved quantities. We therefore linearize about the pair of points $(K_{Fx}, K_{Fy}, \alpha K_{Fz})$ and obtain the bound-state spectrum as in Eqs.~\ref{Eq:Elct_enrg} and~\ref{Eq:Hl_enrg} which now extend along the $z$-direction. The eigenvalues are obtained via the replacement $K_{Fx} \rightarrow K_{Fz}.$

We comment on the effects of unequal coupling of the two orbitals in either lead to the C part. In this case, both symmetric and anti-symmetric linear combinations of the two orbitals (Appendix~\ref{App:st3_N_st3}), couple to the C part, in contrast to the solutions presented here. As shown by numerical results presented in Appendix~\ref{App:Unql}, unequal couplings induce a gap, as well as a spectrum which depends on the global phase difference $\phi$. The first can be understood via a breaking of the orbital-exchange symmetry, while the second can be confirmed via a direct calculation of the Josephson current, as discussed in Appendix~\ref{App:Jsph}.

Finally, we note that the surprising behavior for $s\tau_{3}$ junctions coupled to a single orbital in the C part differs dramatically from that of junctions consisting of leads in a pairing state which has trivial orbital structure. We consider junctions with such a state, which we call $s\tau_{0}$, where the two orbitals have identical pairing functions, including the signs. For degenerate orbitals, we can apply the same analysis as in the $s\tau_{3}-N-s\tau_{3}$ case (Appendix~\ref{App:st3_N_st3}) and we again find that the symmetric linear combination of the two orbitals in either leads couples across the junction. However, the anti-symmetric linear combination is entirely decoupled along the entire junction, and in contrast to $s\tau_{3}$ junctions, imposes no additional constraints on the bound-state solutions. Thus $s\tau_{0}$ junctions for degenerate orbitals behave essentially like conventional single-channel junctions.  This is illustrated in Appendix~\ref{App:st0}. 

\subsection{Non-degenerate orbitals}    

So far, we have focused on the special case of degenerate orbitals for a $s\tau_{3}-N-s\tau_{3}$ junction along $x$. We now consider more realistic cases, where the L and R leads include terms which lift the degeneracy between the two orbitals. As discussed in Sec.~\ref{Sec:Mdls_st3_x}, we consider two types of symmetry-allowed terms which can lift the orbital degeneracy: (i) intra-orbital hybridization terms corresponding to $\xi_{3} \tau_{3}$ in the non-pairing part of the bulk lead Hamiltonians (Eq.~\ref{Eq:BdG_st3_ld}), and (ii) inter-orbital hybridization terms corresponding $\xi_{1} \tau_{1}$ in the same model. We examine the effects of each of these terms on the bound state spectrum separately. 

\begin{figure}[t!]
\includegraphics[width=1.0\columnwidth]
{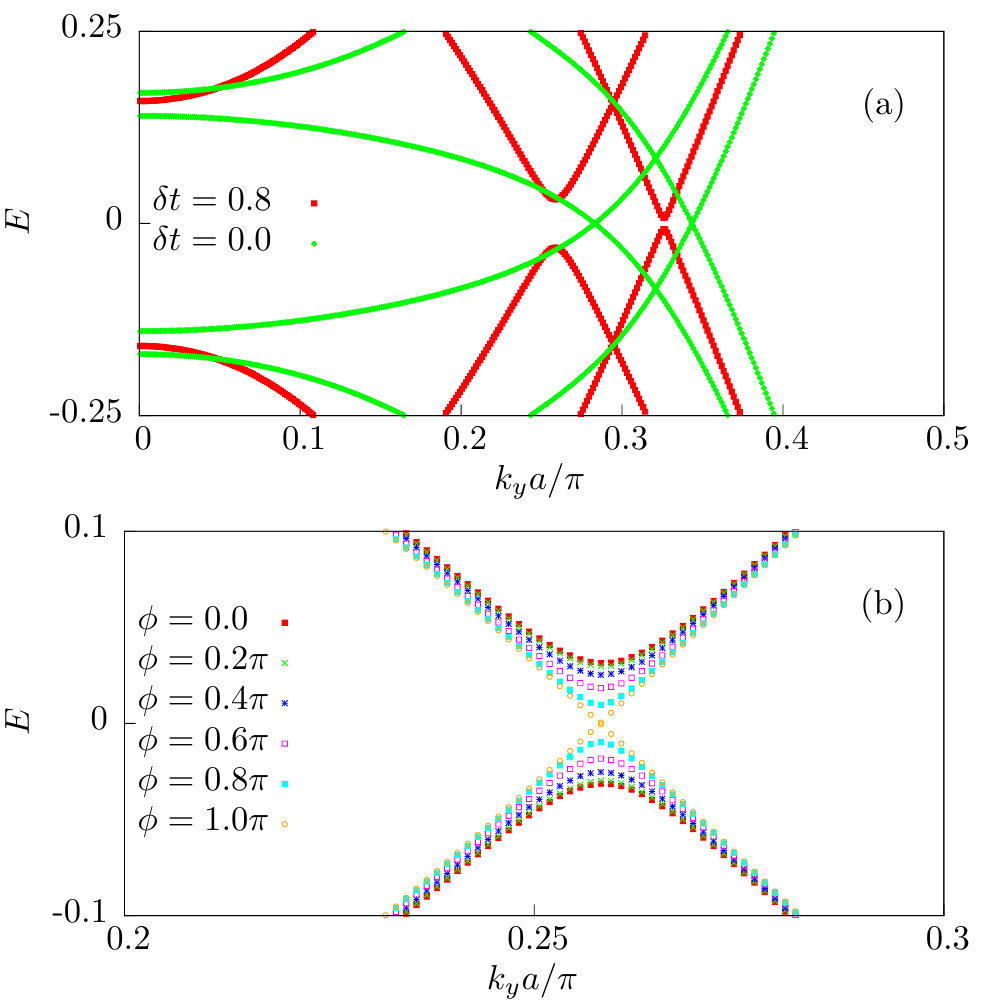}
\caption{The effects of intra-orbital hybridization terms corresponding to $\xi_{3}$ in the bulk of the L/R leads (Eq.~\ref{Eq:BdG_st3_ld}) on the bound state spectrum for a $s\tau_{3}-N-s\tau_{3}$ junction along $x$. We introduce these terms by fixing $t_{1}=1$ and allowing $\delta t= t_{1}- t_{2}$ to vary (Eq.~\ref{Eq:xi3}). The inter-orbital hybridization corresponding to $\xi_{1}$ in Eq.~\ref{Eq:BdG_st3_ld} is set to zero. All other parameters are the same as in Fig.~\ref{Fig:Drft_egnv_mu_vr}~(a) and $\phi=0$. The two orbitals couple identically to the C part. (a) A gap develops as intra-orbital hybridization terms are introduced. For clarity, we show the case for degenerate orbitals $\delta t=0$ (green rhombi), and a case with extreme intra-orbital hybridization for $\delta t=0.8$ (red squares). Note that a finite gap opens for any $\delta t \neq 0$. Intra-orbital hybridization terms break the orbital-exchange symmetry discussed in Sec.~\ref{Sec:Orbl_exch}, allowing the electron- and hole-like states in Eqs.~\ref{Eq:Spnr_elct} and~\ref{Eq:Spnr_hl} to mix via Andreev scattering.~(b) Effect of a global phase difference $\phi$ across the junction on the bound state spectrum with finite intra-orbital hybridization $\delta t=0.8$. In contrast to the degenerate case shown in Fig.~\ref{Fig:Drft_egnv_mu_vr}~(a), the inclusion of finite intra-orbital hybridization allows the spectrum to change with $\phi$. Note that the states become gapless at $\phi=\pi$, much like in a conventional single-channel junction. The sensitivity of the spectrum with $\phi$ is also confirmed via a calculation of the Josephson current in Appendix~\ref{App:Jsph}.}
\label{Fig:Drft_egnv_xi_3}
\end{figure}

We first consider intra-orbital hybridization terms only by fixing $t_{1}=1$ and allowing $\delta t = t_{1}-t_{2}$ to vary (Eq.~\ref{Eq:xi3}). All other parameters are identical to those in the top panel of Fig.~\ref{Fig:Drft_egnv_mu_vr}. In Fig.~\ref{Fig:Drft_egnv_xi_3}~(a) we show the bound state spectrum for the orbital-degenerate case with $\delta t =0$ together with a case with significant intra-orbital hybridization corresponding to $\delta t= 0.8$. The gapless states for the degenerate orbital case become gapped with the inclusion of intra-orbital hybridization terms. These break the orbital-exchange symmetry and thus mix the electron- and hole-like states of the orbital-degenerate case, in accordance with Sec.~\ref{Sec:St3_dgnr}. In panel (b), we show the spectrum for $\delta t =0.8$ as a function of the global phase difference $\phi$. The eigenvalues change with $\phi$ and we recover gapless states at $\pi$ phase difference, as is the case for the typical single-channel junction. The evolution of the bound state spectrum with $\phi$ is also confirmed via a calculation of the Josephson current in Appendix~\ref{App:Jsph}. 

\begin{figure}[t!]
\includegraphics[width=1.0\columnwidth]
{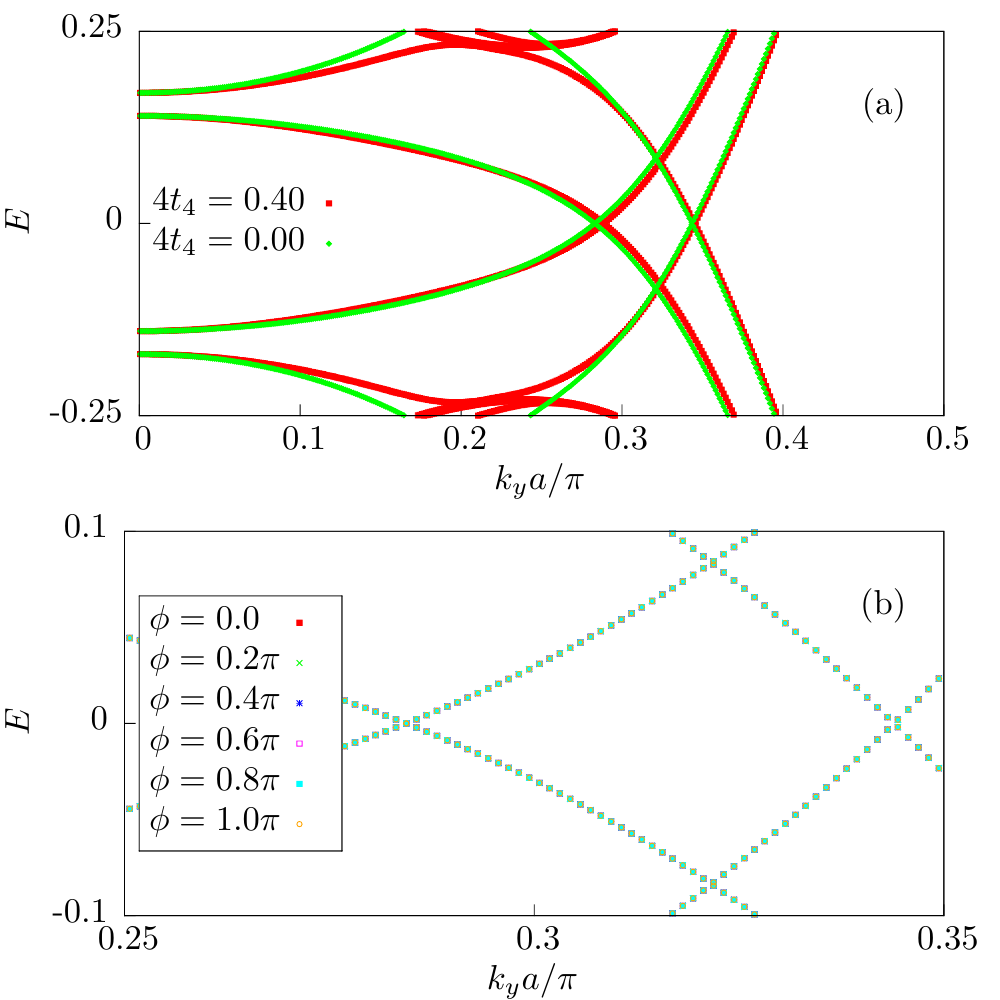}
\caption{The effects of an inter-orbital hybridization term, corresponding to $\xi_{1}$ (Eq.~\ref{Eq:xi1}) in the bulk L and R leads, on the bound state spectrum for a $s\tau_{3}-N-s\tau_{3}$ junction along $x$. The intra-orbital hybridization terms corresponding to $\xi_{3}$ are set to zero. The remaining parameters are the same as in Fig.~\ref{Fig:Drft_egnv_xi_3}.~(a) The bound states remain gapless as $t_{4}$ (Eq.~\ref{Eq:xi1}) increases, in contrast to the intra-orbital hybridization case in Fig.~\ref{Fig:Drft_egnv_xi_3}. The inter-orbital hybridization preserves the orbital-exchange symmetry defined in Sec.~\ref{Sec:Orbl_exch} and thus does not mix the electron- and hole-like states, which remain adiabatically connected with the solutions of the orbital-degenerate junction in Eqs.~\ref{Eq:Spnr_elct} and~\ref{Eq:Spnr_hl}.~(b) The spectrum is insensitive to a global phase difference $\phi$, in contrast to leads with intra-orbital hybridization. These results are also confirmed via a calculation of the Josephson current (Appendix~\ref{App:Jsph}).}
\label{Fig:Drft_egnv_xi_1}
\end{figure}

We now consider the effect of the inter-orbital hybridization terms only, corresponding to $\xi_{1}$ in Eq.~\ref{Eq:BdG_st3_ld}. All other parameters are the same in porevious cases. As shown in Fig.~\ref{Fig:Drft_egnv_xi_1}~(a) the bound state spectrum remains gapless under the inclusion of inter-orbital hybridization, in contrast to intra-orbital hybridization terms. This can also be understood via the orbital-exchange symmetry, since the inter-orbital hybridization terms preserve the former. Similarly, the spectrum is invariant under changes in the global phase difference $\phi$, as indicated in Fig.~\ref{Fig:Drft_egnv_xi_1}~(b). It is then clear that, due to intra-orbital hybridization terms, a general $s\tau_{3}-N-s\tau_{3}$ junction along $x$ exhibits the usual features of a single-channel junction.

\begin{figure}[t!]
\includegraphics[width=1.0\columnwidth]
{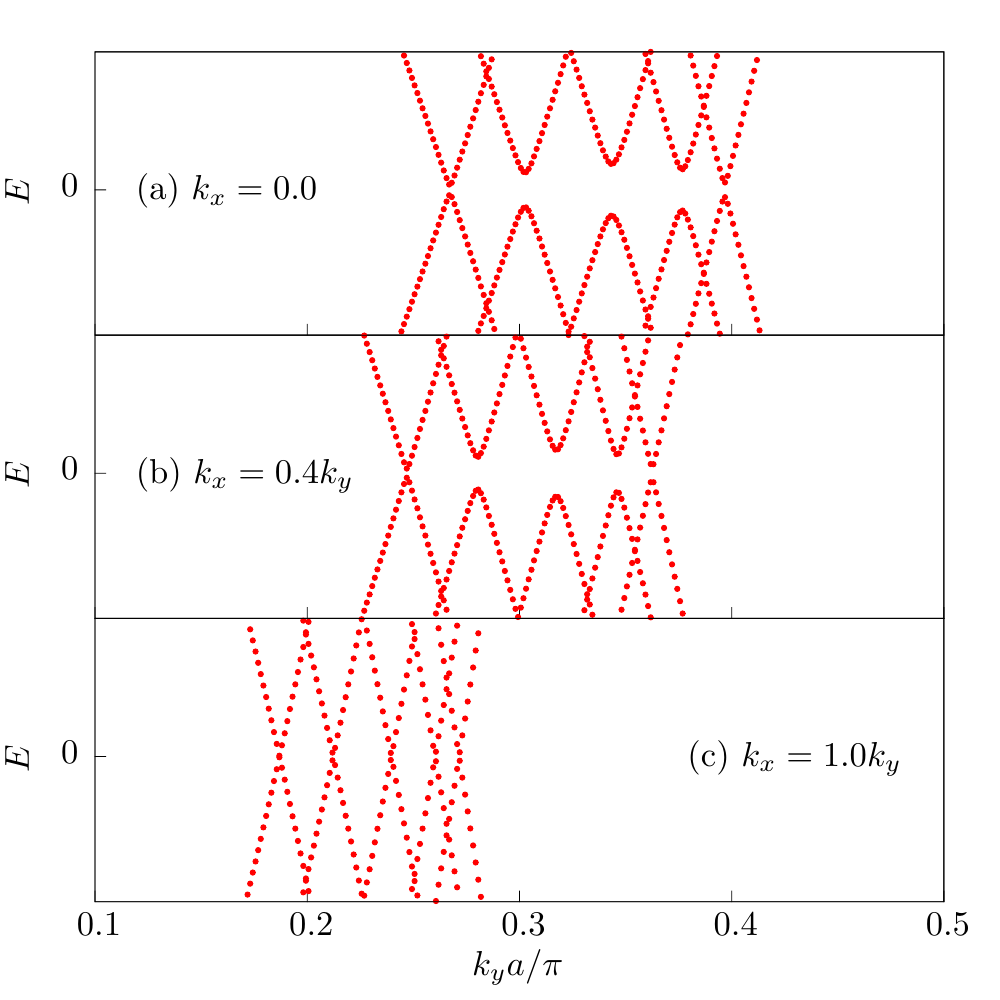}
\caption{Bound state spectrum for a $s\tau_{3}-N-s\tau_{3}$ junction along $z$, calculated from the model in Sec.~\ref{Sec:Mdls_st3_z}, with finite intra- and inter-orbital hybridization. We use $\delta t=0.4$ and $4 t_{4}= 0.4$, $\mu=-3.0$ with a NN $t_{z}=0.2$ along $z$. Note that both $k_{x}$ and $k_{y}$ are conserved in this case. In contrast to the junction along $x$, gapless states are possible here even when intra-orbital hybridization is included. The latter vanishes for $|k_{x}|=|k_{y}|$. For these momenta, we recover the orbital-exchange symmetry defined in Sec.~\ref{Sec:Orbl_exch}.~(a) bound state eigenvalues as a function of $k_{y}$ for fixed $k_{x}=0$. The intra-orbital hybridization terms are finite for these momenta, and the states are gapped, as when the junction is along $x$ (Fig.~\ref{Fig:Drft_egnv_xi_3}.~(b) Same as (a) for $k_{x}=0.4k_{y}$. The gaps shrink but remain finite.~(c) Same as (a) and (b) along the diagonal of the 2D Brillouin zone for $k_{x}=k_{y}$. The gap closes as the model recovers the orbital-exchange symmetry, and the spectrum is similar to that of the junction along  $x$ with inter-orbital hybridization only (Fig.~\ref{Fig:Drft_egnv_xi_1}).}
\label{Fig:Drft_alng_z}
\end{figure}

As a final example, we consider the effects of intra- and inter-orbital hybridization on the junction along $z$. The spectrum in this case can be labeled by both $k_{x}, k_{y}$ conserved momenta. For general values of the latter, the spectrum is gapped, in analogy to the junction along $x$. However, along the diagonals of the two-dimensional Brillouin zone $|k_{x}|=|k_{y}|$, the intra-orbital hybridization terms vanish, and the bound state states again become gapless. This is illustrated in Fig.~\ref{Fig:Drft_alng_z}, and can be understood via the orbital-exchange symmetry, which is recovered along the diagonals. Thus, in contrast to the junction along $x$, the junction along $z$ can exhibit gapless edge states for a rather general model of $s\tau_{3}$ pairing. 

The bound-state spectra of $s\tau_{3}-N-I$ junctions are very similar to those of $s\tau_{3}-N-s\tau_{3}$ junctions. In the limit of degenerate orbitals, the bound states of the former are also electron- or hole like, and they also become gapless for a set of conserved momenta. The analytical solutions in this limit are discussed in Appendix~\ref{App:st0}. The evolution with either intra- or inter-orbital hybridization is also very similar to that in  $s\tau_{3}-N-s\tau_{3}$ junctions and, as such, will not be discussed here in detail. 

\section{$s\tau_{3}-N-s$ junctions}

\label{Sec:s_trvl}

We consider a $s\tau_{3}-N-s$ junction, where the L lead is in a $s\tau_{3}$ pairing state, while the R lead is in a single-channel $s$-wave state. As in the case of $s\tau_{3}-N-s\tau_{3}$ junctions, both orbitals on the L side couple identically to a single orbital in the C metallic part, which in turn couples to a single orbital in the trivially-paired R lead. The model is described in Sec.~\ref{Sec:Mdls_st3_x}. 

We first solve the system in the limit of degenerate orbitals in the L leads, which couple identically to the single C orbital.  The analytical solution is presented in detail in Appendix~\ref{App:st3_N_s}. Here, we summarize some of the most important results. In contrast to the $s\tau_{3}-N-s\tau_{3}$ case, only the L-C interface is subject to both open and continuity BC's. 
Furthermore, as explained in Sec.~\ref{Sec:Orbl_exch}, this model does not preserve the orbital-exchange symmetry, due to the presence of a single channel in the R lead, even when the two orbitals are degenerate. Thus, in contrast to the $s\tau_{3}-N-s\tau_{3}$ case, the $s\tau_{3}-N-s$ junction allows for Andreev scattering mixing electron and hole-like states, and leading to bound states which are generally gapped. However, in the degenerate orbital limit, the bound states states of the $s\tau_{3}-N-s$ are invariant under a change in the global relative phase $\phi$. 

\begin{figure}[t!]
\includegraphics[width=1.0\columnwidth]
{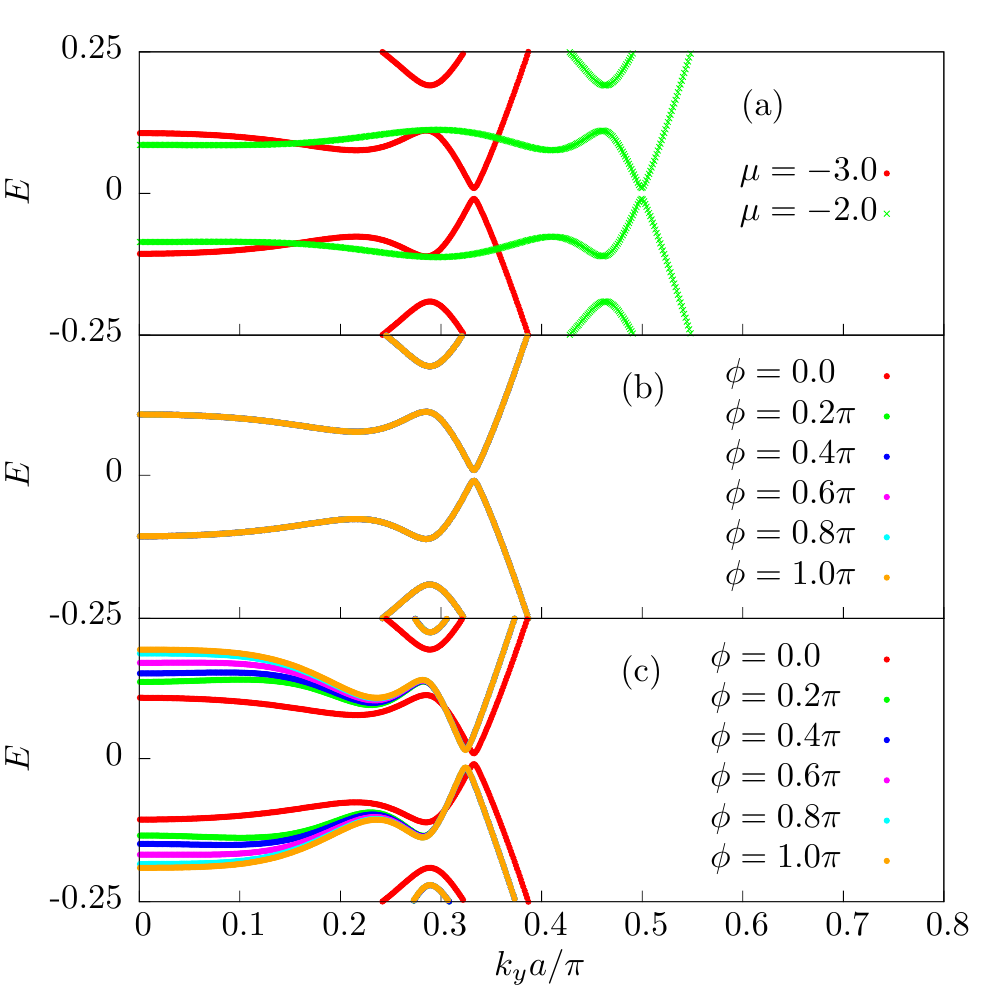}
\caption{Bound state spectrum for a $s\tau_{3}-N-s$ junction along $x$. In all cases, the two orbitals in the L lead couple identically to the single C orbital. All other parameters, unless explicitly stated, are the same as in Fig.~\ref{Fig:Drft_egnv_mu_vr}.~(a) Evolution of the spectrum for degenerate orbitals in the L lead with chemical potential $\mu$. Note that the states are gapped for both values shown here, as well as for any $\mu$. This stands in contrast to the $s\tau_{3}-N-s\tau_{3}$ junction, as shown in Fig.~\ref{Fig:Drft_egnv_mu_vr}. The difference can be attributed to the orbital-exchange symmetry, which is broken for the $s\tau_{3}-N-s$, thus allowing Andreev scattering which mixes electron- and hole-like states.~(b) The spectrum for the degenerate-orbital case of panel (a) with $\mu=-3.0$ is invariant under a change in global relative phase $\phi$. This is due to the absence of intra-orbital hybridization terms, as is also the case with the $s\tau_{3}-N-s\tau_{3}$ junctions in Fig.~\ref{Fig:Drft_egnv_mu_vr}~(a).~(c) The spectrum does change with $\phi$ once intra-orbital hybridization terms are introduced for the L lead, in a manner analogous to Fig.~\ref{Fig:Drft_egnv_xi_3}. This is also confirmed by the calculation of the Josephson current in Appendix~\ref{App:Jsph}.}
\label{Fig:Drft_trvl_s}
\end{figure}

The analytical results are confirmed by the numerical solutions. In Fig.~\ref{Fig:Drft_trvl_s}~(a), we show the bound state spectrum for a $s\tau_{3}-N-s$ junction, whith parameters similar to those of Fig.~\ref{Fig:Drft_egnv_mu_vr}, for two values of $\mu$. We see the presence of a gap in both instances, a gap which also occurs for any value of $\mu$. An analysis of the eigenstates, not shown here for brevity, likewise indicates that the C spinors involve a mixture of electron- and hole-like states. 
Remarkably, the spectrum is invariant under a change in global phase difference $\phi$ in the degenerate-orbital limit, as shown in Fig.~\ref{Fig:Drft_trvl_s}~(b). The spectrum does change with $\phi$ once intra-orbital hybridization terms are introduced, as in the case of $s\tau_{3}-N-s\tau_{3}$ junctions. The bound-state variation with $\phi$ is also confirmed via a calculation of the Josephson current in Appendix~\ref{App:Jsph}. 

\section{Summary and Discussion}

We now summarize our main results and subsequently discuss possible experimental realizations.

\label{Sec:Dscs}

\subsection{Summary of results}

We studied Josephson junctions where the left lead is in an unconventional $s\tau_{3}$ gapped pairing state which involves two orbitals with opposite-sign pairing. Similar pairing states were advanced as promising candidates in alkaline Fe- selenides~\cite{Nica_Yu}, where $d_{xz}, d_{yz}$ orbitals provided the two-dimensional manifold for $s\tau_{3}$, as well as in the heavy-fermion superconductor CeCu$_{2}$Si$_{2}$~\cite{Nica_Si_npj_2021}. We considered junctions where both orbitals of the left lead couple to a single orbital in an intermediate  metallic part. Such arrangements, although unusual, allow for a richer phenomenology when compared to the more typical junctions without cross-coupling. We considered three different types of junctions, referred to as $s\tau_{3}-N-s\tau_{3}$, $s\tau_{3}-N-I$, and $s\tau_{3}-N-s$, where the right lead is in a two-orbital $s\tau_{3}$ pairing, insulating, and trivial  single-orbital $s$-wave pairing states, respectively. We discussed both two-dimensional arrangements with junctions along the $x$-axis, as well as junctions along $z$. 

We studied the three types of junctions in the important limit where the two orbitals of the $s\tau_{3}$ pairing state, in the left and, when appropriate, right leads, are degenerate and couple identically to a single orbital of the central part. Our most striking results can be grouped under two headings. One is the emergence of purely electron and hole-like bound states, which become gapless and degenerate for a set of conserved momenta. The other is that the bound state spectrum is invariant under a change in the global phase difference between left and right superconducting leads. In both aspects, the junctions differ sharply from the typical single- or multi-channel junctions without cross-coupling. The first of these two effects occurs for $s\tau_{3}-N-s\tau_{3}$ and $s\tau_{3}-N-I$ junctions along $x$. Here, the bound states differ from to the typical single-channel junction, where Andreev scattering mixes purely electron- and hole-like states. The absence of Andreev scattering in these cases, which leads to gapless electron- and hole-like states, is due to the combined effects of non-trivial orbital structure of $s\tau_{3}$ pairing and coupling to a single C orbital. We also find that these gapless bound states are protected by an orbital-exchange symmetry. 
However, we stress that this protection is not due to any topological property of the junction, unlike the well-known gapless states which occur in a single-channel $\pi$ junction. The second property, the invariance of the bound state spectrum with global phase difference, manifests in $s\tau_{3}-N-s\tau_{3}$ and $s\tau_{3}-N-s$ junctions along $x$. The invariance of the bound state spectrum points toward a vanishing Josephson current~\cite{Sauls_2018}. We confirmed that this is the case by calculating the Josephson current directly in the tunneling limit. We stress that $s\tau_{3}-N-s\tau_{3}$ junctions in the limit of degenerate orbitals differ dramatically from $s\tau_{0}-N-s\tau_{0}$ junctions in the same limit. The latter involve orbitally-trivial pairing states which are identical to both orbitals, including the sign. In clear contrast to the $s\tau_{3}-N-s\tau_{3}$ junctions, the $s\tau_{0}-N-s\tau_{0}$ bound state spectrum is essentially that of a typical single-channel junction. 

We also considered deviations from the degenerate-orbital limit by introducing intra- and inter-orbital hybridization terms for the normal part of the leads. These terms are allowed by tetragonal symmetry for $d_{xz}, d_{yz}$ orbitals in the context of alkaline Fe-selenides~\cite{Raghu}. We found that intra-orbital hybridization terms break the orbital-exchange symmetry and lead to Andreev scattering which mixes electron and hole-like states, gapping the bound state spectrum of $s\tau_{3}-N-s\tau_{3}$ and $s\tau_{3}-N-I$ junctions along $x$. By contrast, inter-orbital hybridization terms do not break the orbital-exchange symmetry and preserve the gapless electron- and hole-like states in the cases mentioned above. The bound states thus behave as in the intra-orbital hybridization case when both the former and inter-orbital hybridization are present. Important exceptions can occur as exemplified by $s\tau_{3}-N-s\tau_{3}$ and $s\tau_{3}-N-I$ junctions along $z$, with dominant NN hopping along $z$. In this case, the bound states become gapless at the zeroes of the intra-orbital hybridization terms, along the diagonal of the two-dimensional Brillouin zone in our case, since the junction recovers orbital-exchange symmetry at these points.  

We also considered the effects of lifting the orbital degeneracy in the bulk of the leads when a finite global phase difference across the junction is present. Here again we find that the bound state spectrum becomes sensitive to the phase when intra-orbital hybridization terms are in effect for both $s\tau_{3}-N-s\tau_{3}$ and $s\tau_{3}-N-s$ junctions. This is also confirmed by calculations of the Josephson current. In contrast, under the addition of inter-orbital hybridization terms only, the bound state spectrum remains invariant, also in accordance with a vanishing Josephson current.  

The two crucial aspects behind the striking behavior of all of the junctions considered here involve two-orbital matrix-pairing $s\tau_{3}$ in the leads, together with identical coupling of both of these orbitals to a single orbital in the central metallic parts. These setups ensure that the two orbital sectors of $s\tau_{3}$ pairing are entangled in a non-trivial way near the interface with the metallic part, even when the two sectors are decoupled far into the bulk of the leads. This is to be contrasted with cases where the pairing has trivial matrix structure, as exemplified by $s\tau_{0}$ pairing. In those cases, the trivial pairing matrix structure means that two orbitals which are decoupled in the bulk of the leads remain so near the interfaces with the metallic part. Likewise, $s\tau_{3}$ junctions with central metallic regions which likewise involve two orbitals, which couple in a "one-to-one" manner with those of $s\tau_{3}$ would fail to fully capture the effects of the non-trivial matrix structure of $s\tau_{3}$. As exemplified by $s\tau_{3}-N-I$ junctions, the interplay between non-trivial interfaces and non-trivial matrix-pairing suggests that exotic edge states could also be engineered for materials which likely involve $s\tau_{3}$ pairing, such as the alkaline Fe-selenides.

\subsection{Experimental signatures}

Having given a summary of our main results, we now discuss their potential experimental signatures. We stress that, in the limit of degenerate orbitals, junctions involving $s\tau_{3}$ pairing are dramatically different from the typical single-channel junction as well as from junctions involving pairing states with trivial orbital structure, such as the $s\tau_{0}$ states. However, any realizations of the $s\tau_{3}$ junctions proposed here will involve deviations from the ideal case. We argue that such deviations can be made small, in the sense discussed below, allowing a partial observation of the striking properties of the ideal case. 

We consider three likely deviations from the ideal case in the form of: (i) unequal coupling of the two orbitals to the single-orbital of the metallic part, (ii) lifting of the degeneracy of the two orbitals in the leads via symmetry-allowed intra- and inter-orbital hybridization, and (iii) weak disorder in the junction. For (i), unequal coupling to the C part gaps the electron- and hole-like states and
{\it induces a nonzero Josephson effect,}
 even when the orbitals of the leads are degenerate. A single-orbital intermediate region amounts to an effectively single-band system which remains non-superconducting for temperatures above the critical temperature of $s\tau_{3}$, which can be estimated from the transition temperature in alkaline Fe-selenides~\cite{Lee_Science_2017}. The condition of equal coupling to the center part is unlikely to occur for junctions along the in-plane axes of the tetragonal Fe-selenides i.e. along $x$ or $y$, since $d_{xz}$ and $d_{yz}$ orbitals typically cannot couple identically to any other in-plane orbital. However, junctions along $z$ provide better candidates in this context, as the the lobes of $d_{xz}$ and $d_{yz}$ are likely to have comparable overlap with such orbitals as $p_{z}$, $d_{z^{2}}$, and $d_{xy}$ along the $z$ axis. 

A second deviation from the ideal case occurs due to the presence of both intra- and inter-orbital hybridization terms in the leads for junctions involving alkaline Fe-selenide leads. As we have shown, such terms will lead to gapping of the electron- and hole-like states and will furthermore ensure that a static Josephson effect is present. Note however, that, as proposed in Refs.~\cite{Nica_Yu, Nica_Si_npj_2021}, and as reiterated in our text, $s\tau_{3}$ pairing induces a full gap in the bulk of the leads when the pairing amplitude exceeds the band splitting near the FS, which is governed by the intra- and inter-orbital hybridization terms. We expect that this limits the effects of band splitting in the case of the junctions, such that the induced gap for the bound states will be small compared to the bulk gap in the leads. Similarly, the Josephson current, which depends on the strength of the intra-orbital hybridization terms, will likely be finite but strongly suppressed, as compared to a similar setup for pairing states with trivial orbital structure. However, it should be borne in mind that exceptions to this behavior can occur for junctions along $z$. The intra-orbital hybridization terms can vanish by symmetry along certain directions in the two-dimensional Brillouin zone. Here, the junction recovers the orbital-exchange symmetry and the gapless bound states. We do expect a finite Josephson current in these cases, due to the contribution of bulk states away from these points. 

The third source of deviation from the ideal case is the presence of disorder in the junction. The gapless bound states in the ideal degenerate cases are not due to any non-trivial topology, although they are protected by the orbital-exchange symmetry. Therefore, these states are not robust against disorder.   

Due to these inherent limitations, the junctions discussed here cannot singlehandedly indicate $s\tau_{3}$ pairing, but they can nonetheless provide strong supporting evidence. 
\section{Acknowledgements}
OE acknowledge support from National Science Foundation Awards No. DMR 1904716. EMN is supported by ASU startup grant. Work at Rice has been supported by the U.S. Department of Energy, Office of Science, Basic Energy Sciences, under Award 
No. DE-SC0018197 and the Robert A. Welch Foundation under Grant No. C-1411 (Q.S.). One of us (Q.S.) acknowledges
the hospitality of the Aspen Center for Physics, which is supported by NSF grant No. PHY-1607611.

\appendix

 \section{Analytical solutions of the $s\tau_{3}-N-s\tau_{3}$ junction with degenerate orbitals }

\label{App:st3_N_st3}

We consider a $s\tau_{3}-N-s\tau_{3}$ junction with degenerate orbitals/bands in the normal states corresponding to $t_{1}=t_{2}, t_{4}=0$ in Eqs.~\ref{Eq:t1},~\ref{Eq:xi1} together with identical coupling to the single channel of the C part corresponding to $V_{1}=V_{2}$ in Eq.~\ref{Eq:HLC}. In this limit, only the symmetric linear combination of the two orbitals at either L-C and C-R interfaces couples to the single orbital of the C part. 

We apply a local unitary transformation to the L lead

\enni \begin{align}
\begin{pmatrix}
c_{\mathbf{r}, I, \up} 
\\
c_{\mathbf{r}, II, \up}
\\
c^{\dag}_{\mathbf{r}, I, \dn}
\\
c^{\dag}_{\mathbf{r}, II, \dn}
\end{pmatrix}
= \frac{1}{\sqrt{2}}
\begin{pmatrix}
1 & 1 & 0 & 0 
\\
1 & -1 & 0 & 0 
\\
0 & 0 & 1 & 1 
\\
0 & 0 & 1 & -1
\end{pmatrix}
\begin{pmatrix}
c_{\mathbf{r}, 1, \up} 
\\
c_{\mathbf{r}, 2, \up}
\\
c^{\dag}_{\mathbf{r}, 1, \dn}
\\
c^{\dag}_{\mathbf{r}, 2, \dn}
\end{pmatrix}.
\label{Eq:Untr_trns}
\end{align}

\enni Under this transformation, only channel $I$, corresponding to the symmetric linear combination of the operators for the two orbitals, couples to the single orbital of the C part. Furthermore, the transformation leaves the orbital-diagonal tight-binding part of the Hamiltonian invariant (corresponding to $\xi_{0}$ in Eq.~\ref{Eq:BdG_st3_ld}) while it transforms the $s\tau_{3}$ pairing as

\enni \begin{align} 
H_{\text{L, Pair}}= & \sum_{\mathbf{r}}  \Delta 
\left( 
c^{\dag}_{\mathbf{r}, 1 \uparrow} c^{\dag}_{\mathbf{r}, 1 \downarrow}
- 
c^{\dag}_{\mathbf{r}, 2 \uparrow} c^{\dag}_{\mathbf{r}, 2 \downarrow}
\right)
 + \text{H.c.} 
 \notag \\
  & \rightarrow 
 \sum_{\mathbf{r}}  \Delta 
\left( 
c^{\dag}_{\mathbf{r}, I \uparrow} c^{\dag}_{\mathbf{r}, II \downarrow}
- 
c^{\dag}_{\mathbf{r}, II \uparrow} c^{\dag}_{\mathbf{r}, I \downarrow}
\right) \text{+H.c.},
\end{align}

\enni where for simplicity we only consider one spin sector. The model thus reduces to solving for channel $I$ as in a typical, single-channel Josephson junction, while channel $II$ must obey open BC's at the L-C interface. 

We proceed to find solutions of the model in the transformed basis which lie below the bulk gap, corresponding to solutions which decay into the L lead. We first apply a Fourier transform along $y$ and linearize the BdG equations in the vicinity of two points $(\alpha K_{Fx} + q_{x}, K_{Fy} + q_{y})$ on the FS, with $q_{x}, q_{y}$ small, and $\alpha = \pm 1$. This is in analogy to the typical single-channel solution~\cite{Sauls_2018}. After identifying $q_{x} \rightarrow -i \partial_{x}$ in the continuum limit, we separate the BdG ansatz of the L lead into fast- and slowly-varying parts as 

\begin{widetext}

\enni \begin{align}
\begin{pmatrix}
\tilde{u}_{I, k_{y};L} \\
\tilde{u}_{II, k_{y};L} \\
\tilde{v}_{I, k_{y};L} \\
\tilde{v}_{II, k_{y};L}
\end{pmatrix}
= & e^{i K_{Fy}y} e^{i K_{Fx}x} e^{\kappa \left(x + \frac{l}{2} \right) }  
\begin{pmatrix}
u_{\alpha=1, I;L} \\
u_{\alpha=1, II;L} \\
v_{\alpha=1, I;L} \\
v_{\alpha=1, II;L}
\end{pmatrix}
+  e^{i K_{Fy}y} e^{- i K_{F}x} e^{\kappa \left(x + \frac{l}{2} \right)}  
\begin{pmatrix}
u_{\alpha=\bar{1}, I;L} \\
u_{\alpha=\bar{1}, II;L} \\
v_{\alpha=\bar{1}, I;L} \\
v_{\alpha=\bar{1}, II;L}
\end{pmatrix}.
\end{align}

\enni Note that the C part is between $-l/2 \le x  \le l/2$. Since we are looking for solutions which decay into the L lead, we take $\kappa$ to be real and positive. Also note that we consider an ansatz which is a linear superposition of solutions with opposite momenta along $x$. This is due to the BdG coefficients of channel $II$ which must obey open BC at the L-C junction, while the momentum along $y$ is conserved. We ignore corrections due to $q_{y}a$. 

The slowly-varying parts obey the BdG equation 

\enni \begin{align}
\begin{pmatrix}
\alpha \left[ -i v_{Fx} \kappa  \right] - \epsilon & 0 & 0 & \Delta 
\\
0 & \alpha \left[ -i v_{Fx} \kappa  \right] - \epsilon & \Delta & 0 
\\
0 & \Delta^{*} & - \alpha \left[ -i v_{Fx} \kappa  \right] - \epsilon & 0 
\\
\Delta^{*} & 0 & 0 & - \alpha \left[ -i v_{Fx} \kappa  \right] - \epsilon 
\end{pmatrix} 
 \begin{pmatrix}
u_{\alpha, I} \\
u_{\alpha, II} \\
v_{\alpha, I} \\
v_{\alpha, II}
\end{pmatrix}
= & 0,
\end{align}

\end{widetext}

\enni where we introduced the Fermi velocities along $x$ via 

\enni \begin{align}
v_{Fx} = & 2t_{1} \sin(K_{Fx}a).
\end{align}

\enni Note that the first and fourth, and second and third rows respectively decouple in the bulk Hamiltonian and can be solved independently. We find that the solutions in the bulk of the L lead are

\enni \begin{equation}
\begin{pmatrix}
u_{\alpha, I} \\
u_{\alpha, II} \\
v_{\alpha, I} \\
v_{\alpha, II}
\end{pmatrix} 
= 
\begin{pmatrix}
A_{\alpha} \left(\epsilon - i \alpha \Lambda  \right)
\\
B_{\alpha} \left(\epsilon - i \alpha \Lambda  \right)
\\
B_{\alpha} \Delta^{*}
\\
A_{\alpha} \Delta^{*}
\end{pmatrix}
\end{equation}

\enni where $A_{\alpha}, B_{\alpha}$ are coefficients to be determined from the BC's and 

\enni \begin{align}
\kappa = & \frac{\Lambda}{v_{Fx}} 
\\
\Lambda = & \sqrt{\Delta^{2}- \epsilon^{2}},~\text{for}~\epsilon < |\Delta|.
\end{align}

The general solution in the C part, extending from $-l/2 \le x \le l/2$ can be similarly determined to be of the form 

\enni \begin{align}
\begin{pmatrix}
\tilde{u}_{k_{y}; C} \\
\tilde{v}_{k_{y}; C} \\
\end{pmatrix}
= & e^{i K_{F}x} 
\begin{pmatrix}
E_{1} e^{i \frac{\epsilon x}{v_{Fx}}} \\
G_{1} e^{-i \frac{\epsilon x}{v_{Fx}}}
\end{pmatrix}
+  e^{-i K_{F}x} 
\begin{pmatrix}
E_{\bar{1}} e^{-i \frac{\epsilon x}{v_{Fx}}} \\
G_{\bar{1}} e^{i \frac{\epsilon x}{v_{Fx}}}
\end{pmatrix}.
\end{align}

\enni The solutions in the R lead, which decay away from the C-R interface for $x >l/2$ are

\begin{widetext}

\enni \begin{align}
\begin{pmatrix}
\tilde{u}_{I, k_{y}; R} \\
\tilde{u}_{II, k_{y}; R} \\
\tilde{v}_{I, k_{y}; R} \\
\tilde{v}_{II, k_{y}; R}
\end{pmatrix}
= & e^{i K_{F}x} e^{-\kappa \left(x - \frac{l}{2} \right) }  
\begin{pmatrix}
M_{1} (\epsilon + i \Lambda) \\
N_{1} (\epsilon + i \Lambda) \\
N_{1} |\Delta| e^{-i\phi} \\
M_{1} |\Delta| e^{-i\phi}
\end{pmatrix}
+  e^{-i K_{F}x} e^{-\kappa \left(x - \frac{l}{2} \right) }  
\begin{pmatrix}
M_{\bar{1}} (\epsilon - i \Lambda) \\
N_{\bar{1}} (\epsilon - i \Lambda) \\
N_{\bar{1}} |\Delta| e^{-i\phi} \\
M_{\bar{1}} |\Delta| e^{-i\phi}
\end{pmatrix}.
\end{align}

\end{widetext}

\enni Note that we have introduced a global phase $\phi$ in the pairing of the R lead. 

For our purposes, it proves convenient to parameterize all of the coefficients in terms of an overall phase and a relative phase as in 

\enni \begin{align}
A_{\alpha} = & |A| e^{i \theta^{0}_{A}} e^{i \alpha \theta_{A}},
\end{align}

\enni and similarly for all $\alpha$-dependent quantities. We also parameterize the factors which enter in the general solutions for the leads as

\enni \begin{align}
\epsilon \pm i \alpha \Lambda = & |\Delta| e^{\pm i \alpha \theta} 
\\
\theta = & \arg\left(\epsilon + i  \Lambda \right).
\end{align}

We now consider the BC's. As discussed previously, channel $II$ in the L and R leads does not couple to the C part, and thus must obey open BC's at the L-C and C-R interfaces, respectively:

\enni \begin{align}
\tilde{u}_{II, k_{y};L}\left(x = \frac{-l}{2} \right) = & 0 \\
\tilde{v}_{II, k_{y};L}\left(x = \frac{-l}{2} \right) = & 0 
\\
\tilde{u}_{II, k_{y};R}\left(x = \frac{l}{2} \right) = & 0 \\
\tilde{v}_{II, k_{y};R}\left(x = \frac{l}{2} \right) = & 0.
\end{align}

\enni By contrast, channel $I$ couples across the junction and therefore satisfies the continuity conditions 

\enni \begin{align}
\tilde{u}_{I, k_{y};L}\left(x = \frac{-l}{2} \right) = & \tilde{u}_{k_{y}; C}\left(x = \frac{-l}{2} \right) \label{Eq:Cntn_1} \\
\tilde{v}_{I, k_{y};L}\left(x = \frac{-l}{2} \right) = &  \tilde{v}_{k_{y}; C}\left(x = \frac{-l}{2} \right) \label{Eq:Cntn_2}
\\
\tilde{u}_{I, k_{y};R}\left(x = \frac{l}{2} \right) = & \tilde{u}_{k_{y}; C}\left(x = \frac{l}{2} \right) \\
\tilde{v}_{I, k_{y};R}\left(x = \frac{l}{2} \right) = &  \tilde{v}_{k_{y}; C}\left(x = \frac{l}{2} \right) 
\end{align}

\enni Expressed in term of the $\alpha$- dependent quantities these amount to 

\enni \begin{align}
- \frac{K_{Fx}l}{2} + \theta_{B} - \theta = & \frac{\pi}{2} + b \pi 
\\
- \frac{K_{Fx}l}{2} + \theta_{A} = & \frac{\pi}{2} + a \pi
\\
\frac{K_{Fx}l}{2} + \theta_{N} + \theta = & \frac{\pi}{2} + n \pi 
\\
\frac{K_{Fx}l}{2} + \theta_{M} = & \frac{\pi}{2} + m \pi, 
\end{align}

\enni for the open BC's. $a, b, m$, and $n$ are arbitrary integers. The continuity conditions imply that 

\enni \begin{align}
\theta_{A} - \theta +\frac{\epsilon l}{v_{Fx}} = & \theta_{M}+\theta 
\\
\theta_{B}-\frac{\epsilon l}{v_{Fx}} = & \theta_{N}
\end{align}

\enni \begin{align}
|A| = & |M| \\
|B| = & |N|
\end{align}

\enni \begin{align}
\theta^{0}_{A} = & \theta^{0}_{M} \\
\theta^{0}_{B} = & \theta^{0}_{M} - \phi.
\end{align}

\enni Note that the \emph{overall phase} $\theta^{0}_{B}$ can incorporate the global phase difference of the pairing $\phi$. This is in contrast to the typical single-channel case and it leads to an insensitivity of the bound state spectrum w.r.t. to $\phi$. 

Note that there are five unknowns consisting of the five relative phase $\theta_{A}, \theta_{B}, \theta_{M}, \theta_{N}$ together with the eigenenergies $\epsilon$. However, due to the open BC's, there are six equations. Therefore, non-trivial solutions cannot be found for this system of equations. We consider instead solutions where either $A_{\alpha}, M_{\alpha}$ or $B_{\alpha}, N_{\alpha}$ are trivially zero. In either of these cases, the system involving the relative phase and $\epsilon$ reduces to three equations with three unknowns. Importantly, these solutions are found to be either hole- or electron-like, as either the $u$ or $v$ BdG coefficients vanish in the C part. 

The electron-like solutions obtained in this manner are 

\begin{widetext}

\enni \begin{align}
\begin{pmatrix}
\tilde{u}_{I, k_{y}; L - e} \\
\tilde{u}_{II, k_{y}; L - e} \\
\tilde{v}_{I, k_{y}; L - e} \\
\tilde{v}_{II, k_{y}; L - e}
\end{pmatrix}
= & 2|A| |\Delta| e^{i \theta^{0}_{A}} 
\begin{pmatrix}
\cos \left[ K_{Fx} x  -\frac{\epsilon l}{2 v_{Fx}} + \frac{(a+m+1)\pi} {2} \right] 
\\
0 
\\
0 
\\
\cos \left[ K_{Fx} \left( x + \frac{l}{2} \right)+ \frac{\pi}{2} + a \pi \right] 
\end{pmatrix}.
\end{align}

\enni \begin{align}
\begin{pmatrix}
\tilde{u}_{k_{y}; C - e} \\
\tilde{v}_{k_{y}; C - e} \\
\end{pmatrix}
= & 2|A| |\Delta| e^{i \theta^{0}_{A}} 
\begin{pmatrix}
\cos \left[ K_{Fx} x + \frac{\epsilon x}{v_{Fx}} + \frac{(a+m+1)\pi}{2}  \right] 
\\
0 
\end{pmatrix}.
\end{align}

\enni \begin{align}
\begin{pmatrix}
\tilde{u}_{I, k_{y}; R-e} \\
\tilde{u}_{II, k_{y}; R-e} \\
\tilde{v}_{I, k_{y}; R-e} \\
\tilde{v}_{II, k_{y}; R-e}
\end{pmatrix}
= & 2|A| |\Delta| e^{i \theta^{0}_{A}} 
\begin{pmatrix}
\cos \left[ K_{Fx} x  + \frac{\epsilon l}{2v_{Fx}} + \frac{(a+m+1)\pi}{2} \right] 
\\
0 
\\
0 
\\
\cos \left[ K_{Fx} \left( x - \frac{l}{2} \right)+ \frac{\pi}{2} + m \pi \right] 
\end{pmatrix}.
\end{align}

\enni Using 

\enni \begin{align}
\arctan(\theta) = \frac{\Lambda}{\epsilon}
\end{align}

\enni we determine the eigenvalues 

\enni \begin{align}
\frac{\epsilon}{|\Delta|} = & \pm 
\cos\left( \frac{\epsilon l}{2 v_{Fx}} + \frac{K_{Fx}l}{2} + \frac{(a-m) \pi}{2} \right). 
\end{align}

\enni Similarly, the hole-like solutions are 

\enni \begin{align}
\begin{pmatrix}
\tilde{u}_{I, k_{y}; L - h} \\
\tilde{u}_{II, k_{y}; L - h} \\
\tilde{v}_{I, k_{y}; L - h} \\
\tilde{v}_{II, k_{y}; L - h}
\end{pmatrix}
= & 2|B| e^{i \theta^{B}_{0}} 
\begin{pmatrix}
0
\\
\cos \left[ K_{Fx} \left( x + \frac{l}{2} \right)+ \frac{\pi}{2} + b \pi \right]
\\
\cos \left[ K_{Fx}x+\frac{\epsilon l}{2v_{Fx}} + 
 \frac{\pi}{2}+\frac{(n+b+1)\pi}{2} \right] 
\\
0  
\end{pmatrix}.
\end{align}

\enni \begin{align}
\begin{pmatrix}
\tilde{u}_{k_{y}; C - h} \\
\tilde{v}_{k_{y}; C - h} \\
\end{pmatrix}
= & 2|B| e^{i \theta^{B}_{0}} 
\begin{pmatrix}
0
\\
\cos \left[ K_{Fx} x - \frac{\epsilon x}{v_{Fx}} +\frac{(n+b+1)\pi}{2} \right] 
\end{pmatrix}.
\end{align}

\enni \begin{align}
\begin{pmatrix}
\tilde{u}_{I, k_{y}; R-h} \\
\tilde{u}_{II, k_{y}; R-h} \\
\tilde{v}_{I, k_{y}; R-h} \\
\tilde{v}_{II, k_{y}; R-h}
\end{pmatrix}
= & 2|B| e^{i (\theta^{B}_{0}+ \phi)} 
\begin{pmatrix}
0
\\
\cos \left[ K_{Fx} \left( x - \frac{l}{2} \right)+ \frac{\pi}{2} + n \pi \right]
\\
\cos \left[ K_{Fx}x-\frac{\epsilon l}{2v_{Fx}} +\frac{(n+b+1)\pi}{2} \right] 
\\
0  
\end{pmatrix}.
\end{align}

\enni with eigenvalues

\enni \begin{align}
\frac{\epsilon}{|\Delta|} 
= & \pm 
\cos\left( \frac{\epsilon l}{2 v_{Fx}} - \frac{K_{Fx}l}{2} + \frac{(n-b) \pi}{2} \right). 
\end{align}

\end{widetext}

\enni The solutions in the original basis, obtained via the inverse transformation of Eq.~\ref{Eq:Untr_trns} can be obtained from the solutions shown above. 

\section{$s\tau_{3}-N-s\tau_{3}$ junctions with unequal coupling to the C part}

\label{App:Unql}

\begin{figure}[ht!]
\includegraphics[width=1.0\columnwidth]
{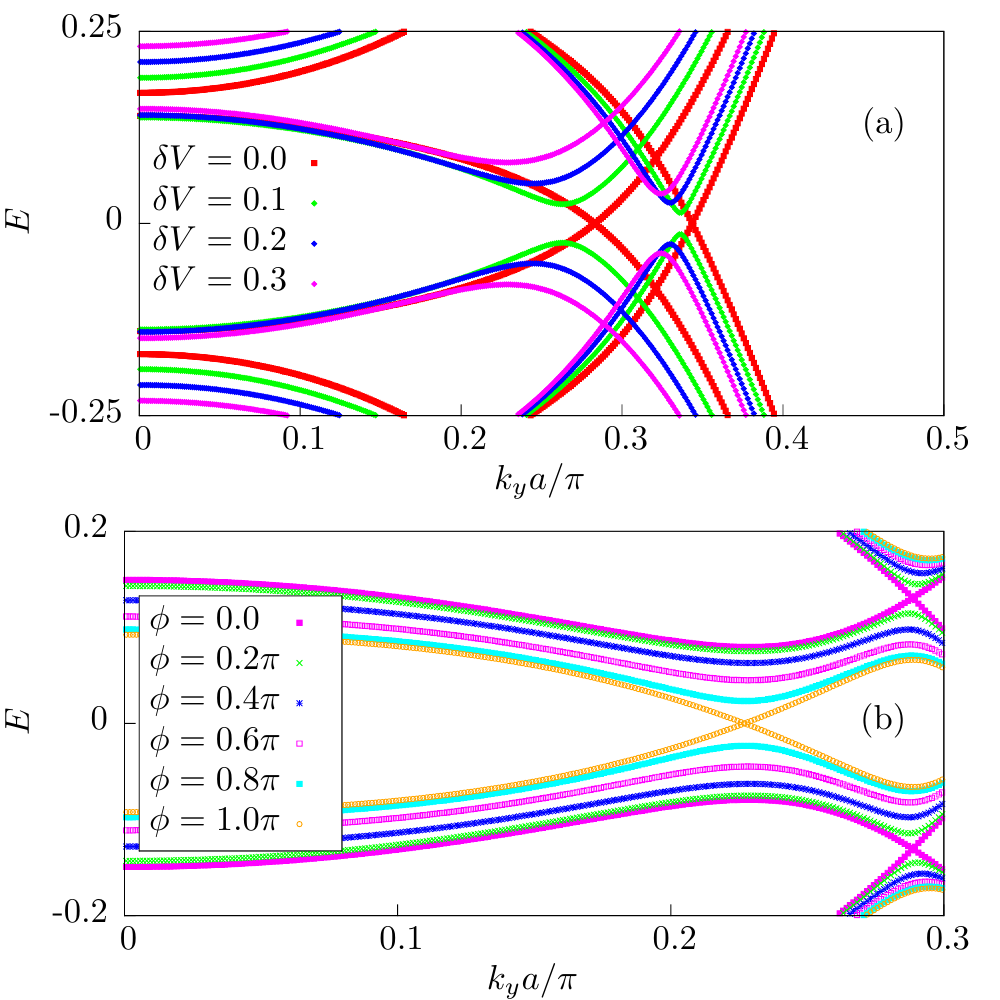}
\caption{Bound state spectrum for a $s\tau_{3}-N-s\tau_{3}$ junction along $x$, where the couplings to the single orbital of the C part, $V_{1}$ and $V_{2}$ (Eq.~\ref{Eq:HLC}), are unequal. We consider degenerate orbitals and compare to the results for $V_{1}=V_{2}$ in Fig.~\ref{Fig:Drft_egnv_mu_vr} ~(a). In contrast to the case with equal $V_{1}=V_{2}$, the bound states acquire a gap with increasing $\delta V = V_{1} - V_{2}$. This can be understood via a broken orbital-exchange symmetry which mixes the purely electron- and hole-like states via Andreev scattering.~(b) The spectrum changes as a function of global phase difference across the junction $\phi$. The results are for $\delta V=0.3$. This is in contrast to the case with $\delta V=0$, which is shown in Fig.~\ref{Fig:Drft_egnv_mu_vr}. }
\label{Fig:Drft_V_vt}
\end{figure}

In Fig.~\ref{Fig:Drft_V_vt}~(a), we show the bound state spectrum for a $s\tau_{3}-N-s\tau_{3}$ junction along $x$ as a function of $\delta V= V_{1}-V_{2}$, the difference between the coupling constants of the two degenerate orbitals in either lead to the single orbital of the C part. The results show that, in clear contrast to the case of equal-coupling $\delta V=0$, a finite $\delta V$ breaks the orbital-exchange symmetry and gaps the purely electron- and hole-like states. In panel (b) we show the spectrum as a function of $\phi$, for fixed $\delta V=0.7$. Again in contrast to the $\delta V=0$ cases, the spectrum becomes dependent on $\phi$, a feature which is also confirmed via a calculation of the Josephson current in Appendix~\ref{App:Jsph}. 

\section{$s\tau_{0}-N-s\tau_{0}$ junctions}

\label{App:st0}

\begin{figure}[t!]
\includegraphics[width=1.0\columnwidth]
{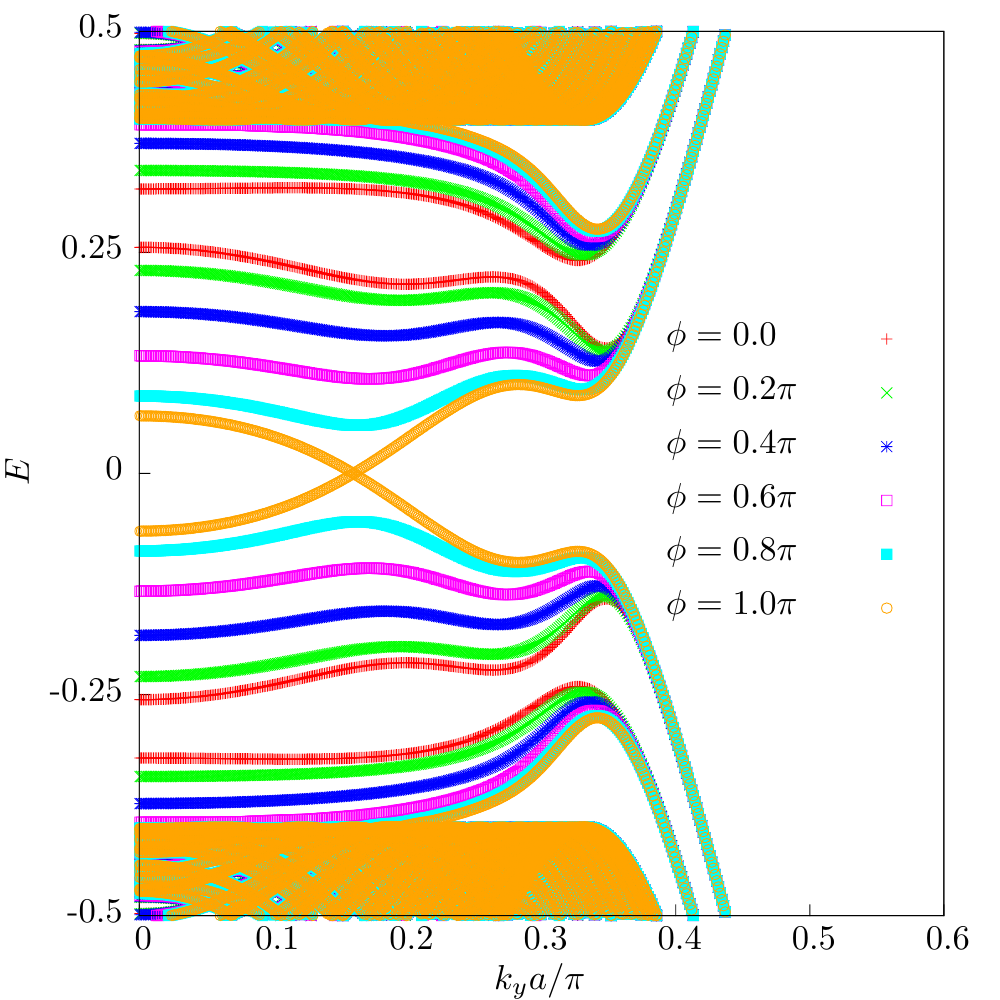}
\caption{Bound state spectrum for a $s\tau_{0}-N-s\tau_{0}$ junction along $x$, where the two orbitals have identical pairing functions, as a function of global phase $\phi$. In contrast to $s\tau_{3}$ pairing, $s\tau_{0}$ is an orbitally-trivial pairing state. The overall setup of the junction is the same as in the $s\tau_{3}$ cases, with each of the two degenerate orbitals in either L and R leads coupling identically to a single orbital in the C part. All of the parameters of the model are the same as in Fig.~\ref{Fig:Drft_egnv_mu_vr}~(a). As indicated by the results, the bound state spectrum for a $s\tau_{0}-N-s\tau_{0}$ junction differs dramatically from a $s\tau_{3}-N-s\tau_{3}$ junction in two main aspects: (i) The spectrum for $\phi=0$ is gapped, and (ii) for varying $\phi$ the spectrum evolves much like a typical single-channel junction, becoming gapless at $\phi=\pi$. In $s\tau_{0}$ junctions, the symmetric linear combination of the two orbitals couples across the junction as in the single-orbital case, while the remaining antisymmetric linear combination decouples entirely.}
\label{Fig:Drft_stau0}
\end{figure}

In this section, we consider $s\tau_{0}-N-s\tau_{0}$ junctions along $x$, where the leads are in $s\tau_{0}$ pairing states. In these cases, the pairing functions for both orbitals are identical. We consider degenerate orbitals which couple to a single orbital in the C part identically. We show that, in clear contrast to $s\tau_{3}-N-s\tau_{3}$ junctions, $s\tau_{0}-N-s\tau_{0}$ junctions behave essentially like a typical single-channel junction. In Fig.~\ref{Fig:Drft_stau0} we illustrate the evolution of the bound state spectrum as a function of global phase difference $\phi$. 

\section{Analytical solutions of the $s\tau_{3}-N-I$ junction with degenerate orbitals}

\label{App:I}

The solution in the continuum limit is very similar to that of the $s\tau_{3}-N-s\tau_{3}$ case. Instead of continuity BC's at the C-R interface, the C spinors obey open BC's. Using the ansatze and the conventions of Appendix~\ref{App:st3_N_st3}, these conditions amount to 

\enni \begin{align}
\frac{K_{Fx} l}{2} + \frac{\epsilon l}{2 v_{Fx}} + \theta_{E} = & \frac{\pi }{2} + p \pi
\\
\frac{K_{Fx} l}{2} - \frac{\epsilon l}{2 v_{Fx}} + \theta_{G} = & \frac{\pi}{2} + r\pi,
\end{align}

\enni where $p, r$ are integers. As for the $s\tau_{3}-N-s\tau_{3}$ junctions, we obtain electron- and hole-like solutions of the form

\enni \begin{align}
& \begin{pmatrix}
u^{L}_{1e} & u^{L}_{2e} & v^{L}_{1e} & v^{L}_{2e}
\end{pmatrix}^{T}
\notag \\
= & 
2|A| e^{i\theta^{A}_{0}} |\Delta|
\begin{pmatrix}
\sin\left[ K_{Fx} \left( x- \frac{l}{2} \right) - \frac{\epsilon l}{v_{Fx}} + p \pi  \right] 
\\
\sin\left[ K_{Fx} \left( x- \frac{l}{2} \right) - \frac{\epsilon l}{v_{Fx}} + p \pi  \right] 
\\
\sin\left[ K_{Fx} \left( x+ \frac{l}{2} \right) + a \pi \right]
\\
- \sin\left[ K_{Fx} \left( x+ \frac{l}{2} \right) + a \pi \right] 
\end{pmatrix}
\end{align}

\enni \begin{align}
& \begin{pmatrix}
u^{C}_{e} v^{C}_{e} 
\end{pmatrix}^{T} 
\notag \\ 
= & 2|A| e^{i\theta_{0}} |\Delta|
\begin{pmatrix}
\sin\left[ 
K_{Fx}\left( 
x -\frac{l}{2}
\right)
+ \frac{\epsilon}{v_{Fx}}
\left( 
x - \frac{l}{2} \right) + p \pi
\right]
\\
0
\end{pmatrix}
\end{align}

\enni with eigenvalues

\enni \begin{align}
\frac{\epsilon}{|\Delta|} = & \pm 
\cos\left( \frac{\epsilon l}{v_{Fx}} +  K_{Fx}l - (p-a) \pi \right) 
\end{align}

\enni for the electron-like states. The hole-like solutions are

\enni \begin{align}
& \begin{pmatrix}
u^{L}_{1h} & u^{L}_{2h} & v^{L}_{1h} & v^{L}_{2h}
\end{pmatrix}^{T}
\notag \\
= & 
2|B| e^{i\theta^{B}_{0}} |\Delta|
\begin{pmatrix}
\sin\left[ 
K_{Fx} 
\left(
x + \frac{l}{2}
\right) + b\pi
\right]
\\
- \sin\left[ 
K_{Fx} 
\left(
x + \frac{l}{2}
\right) + b\pi
\right]
\\
\sin\left[ 
K_{Fx}\left(
x - \frac{l}{2} 
\right)
+ \frac{\epsilon l}{v_{Fx}} + r\pi 
\right]
\\
\sin\left[ 
K_{Fx}\left(
x - \frac{l}{2} 
\right)
+ \frac{\epsilon l}{v_{Fx}} + r\pi 
\right]
\end{pmatrix}
\end{align}

\enni \begin{align}
& \begin{pmatrix}
u^{C}_{h} v^{C}_{h} 
\end{pmatrix}^{T} 
\notag \\ 
= & 2|B| e^{i\theta^{0}_{B}} |\Delta|
\begin{pmatrix}
0
\\
\sin\left[ 
K_{Fx}\left( 
x -\frac{l}{2}
\right)
- \frac{\epsilon}{v_{Fx}}
\left( 
x - \frac{l}{2} \right)+ r \pi
\right]
\end{pmatrix}
\end{align}

\enni with eigenvalues

\enni \begin{align}
\frac{\epsilon}{|\Delta|} = & \pm 
\cos\left(\frac{\epsilon l}{v_{Fx}} -K_{Fx}l +  (r-b) \pi \right). 
\end{align}

\enni As before, $|A|, |B|$ are normalization constants, $\theta^{0}_{A/B}$ are arbitrary phases, while $a,b,p,r$ are arbitrary integers.

 \section{Analytical solutions of the $s\tau_{3}-N-s$ junction with degenerate orbitals}

\label{App:st3_N_s}

Consider a junction of the $s\tau_{3}-N-s$ along $x$, where the R lead is in a single-channel, $s$-wave pairing state. The model for this junction was introduced in Sec.~\ref{Sec:Mdls_st3_x}. Here, we tackle this analytically for the case where the two channels of the L leads, which are in a $s\tau_{3}$ pairing state, are degenerate, and couple identically to the orbital in the C part. 

We proceed along the same lines as the $s\tau_{3}-N-s\tau_{3}$ case in the continuum limit. As in that case, only the symmetric linear combination of the two channels of the L lead couples to the C part. We therefore consider the same ansatz for the BdG coefficients in the L lead which also obey the same open and continuity BC at the L-C interface as before (Eqs.~\ref{Eq:Cntn_1}-\ref{Eq:Cntn_2}). The main distinction is due to the presence of a single channel in the R lead with general solutions in the bulk given by 

\begin{widetext}

\enni \begin{align}
\begin{pmatrix}
\tilde{u}_{k_{y}; R} \\
\tilde{v}_{k_{y}; R}
\end{pmatrix}
= & e^{i K_{F}x} e^{-\kappa \left(x - \frac{l}{2} \right) }  
\begin{pmatrix}
M_{1} (\epsilon + i \Lambda) \\
M_{1} |\Delta| e^{-i\phi}
\end{pmatrix}
+  e^{-i K_{F}x} e^{-\kappa \left(x - \frac{l}{2} \right) }  
\begin{pmatrix}
M_{\bar{1}} (\epsilon - i \Lambda) \\
M_{\bar{1}} |\Delta| e^{-i\phi} \\
\end{pmatrix}.
\end{align}

\end{widetext}

\enni Furthermore, these BdG coefficients obey a single continuity BC at the C-R interface. Adopting the same conventions as in the $s\tau_{3}-N-s\tau_{3}$ case, we summarize the boundary conditions as 

\enni \begin{align}
- \frac{K_{Fx}l}{2} + \theta_{B} - \theta = & \frac{\pi}{2} + b \pi 
\\
- \frac{K_{Fx}l}{2} + \theta_{A} = & \frac{\pi}{2} + a \pi
\end{align}

\enni \begin{align}
\theta_{A} - \theta +\frac{\epsilon l}{v_{Fx}} = & \theta_{M}+\theta 
\\
\theta_{B}-\frac{\epsilon l}{v_{Fx}} = & \theta_{M}
\end{align}

\enni for the relative phases and  

\enni \begin{align}
|A| = & |M| \\
|B| = & |N|
\end{align}

\enni \begin{align}
\theta^{0}_{A} = & \theta^{M}_{0} \\
\theta^{0}_{B} = & \theta^{0}_{M} - \phi.
\end{align}

\enni for the amplitudes and global phases of the BdG coefficients. $a,b$ are arbitrary integers.  

In contrast to the case of the $s\tau_{3}-N-s\tau_{3}$ junction, the absence of additional open BC for the R lead ensures that the equations for the relative phases and $\epsilon$ have a unique solution. After straightforward algebra, we determine the eigenvalues from 

\enni \begin{align}
\frac{\epsilon}{|\Delta|} = & \pm 
\cos\left(  \frac{2\epsilon l}{3 v_{Fx}} + \frac{(a-b)\pi}{3} \right) .
\end{align}

\enni The corresponding states are 

\begin{widetext}

\enni \begin{align}
\begin{pmatrix}
\tilde{u}_{1, k_{y}, L} \\
\tilde{u}_{2, k_{y}, L, e} \\
\tilde{v}_{1, k_{y}, L, e} \\
\tilde{v}_{2, k_{y}, L, e}
\end{pmatrix}
= & 2|A| |\Delta| e^{i \theta^{A}_{0}}
\begin{pmatrix}
 \sin \left[ K_{Fx} \left(x + \frac{l}{2} \right) - \frac{2\epsilon l}{3 v_{Fx}} + \frac{(2a + b) \pi}{3} \right] + e^{- i \phi} \sin \left[ K_{Fx} \left( x + \frac{l}{2} \right) + b \pi \right]
\\
\\
\sin \left[ K_{Fx} \left(x + \frac{l}{2} \right) - \frac{2\epsilon l}{3 v_{Fx}} + \frac{(2a + b) \pi}{3} \right] - e^{- i \phi} \sin \left[ K_{Fx} \left( x + \frac{l}{2} \right) + b \pi \right]  
\\
\\
e^{- i \phi} \sin \left[ K_{Fx} \left(x + \frac{l}{2} \right) + \frac{2\epsilon l}{3 v_{Fx}} + \frac{(a + 2b) \pi}{3} \right] 
+ \sin \left[ K_{Fx} \left(x + \frac{l}{2} \right) + a \pi \right]
\\
\\
e^{- i \phi} \sin \left[ K_{Fx} \left(x + \frac{l}{2} \right)+ \frac{2\epsilon l}{3 v_{Fx}} + \frac{(a + 2b) \pi}{3} \right] 
-\sin \left[ K_{Fx} \left(x + \frac{l}{2} \right) + a \pi \right] 
\end{pmatrix}.
\end{align}

\enni \begin{align}
\begin{pmatrix}
\tilde{u}_{k_{y}, C} \\
\tilde{v}_{k_{y}, C} \\
\end{pmatrix}
= & 2|A| |\Delta| e^{i \theta_{0}} 
\begin{pmatrix}
\sin \left[ K_{Fx} \left( x + \frac{l}{2} \right) +  \frac{\epsilon x}{v_{Fx}}  - \frac{\epsilon l}{6 v_{Fx}} + \frac{(2a + b) \pi}{3} \right] 
\\
\\
e^{-i \phi} \sin \left[ K_{Fx} \left( x + \frac{l}{2} \right) - \frac{\epsilon x}{v_{Fx}}+ \frac{\epsilon l}{6 v_{Fx}} + \frac{(a + 2b) \pi}{3}  \right]  
\end{pmatrix}.
\label{Eq:stau3_N_s_C}
\end{align}

\enni \begin{align}
\begin{pmatrix}
\tilde{u}_{k_{y}, R} \\
\tilde{v}_{k_{y}, R} \\
\end{pmatrix}
= & 2|A| |\Delta| e^{i \theta_{0}} 
\begin{pmatrix}
\sin \left[ K_{Fx} \left( x + \frac{l}{2} \right) +  \frac{\epsilon l}{3 v_{Fx}}  + \frac{(2a + b) \pi}{3} \right] 
\\
\\
e^{-i \phi} \sin \left[ K_{Fx} \left( x + \frac{l}{2} \right) - \frac{\epsilon l}{3v_{Fx}} + \frac{(a + 2b) \pi}{3}  \right]  
\end{pmatrix}.
\end{align}

\end{widetext}

\enni Note that the C spinor is a mixture of electron- and hole-like solutions, as in the case with typical Andreev bound states.

\section{Josephson current in the tunneling limit}

\label{App:Jsph}

In Sec.~\ref{Sec:St3_dgnr} and~\ref{Sec:s_trvl}, we showed that in the limit of degenerate orbitals which couple identically to a single orbital in the C part, $s\tau_{3}-N-s\tau_{3}$ and $s\tau_{3}-N-s$ junctions exhibit bound state spectra which are insensitive to changes in the relative global phase $\phi$. In view of the relation between the Josephson current and the derivative of the ground-state energy with $\phi$, which is typically determined by the Andreev bound state spectrum, our results suggest the absence of a static Josephson effect. 

In order to confirm these findings, we consider simplified models of the $s\tau_{3}-N-s\tau_{3}$ and $s\tau_{3}-N-s$ junctions and determine the Josephson current in the tunneling limit. We follow the standard approach in Ref.~\cite{Sauls_2018}. More precisely, we introduce a Hamiltonian which includes the leads but which also involves direct tunneling between the latter:

\enni \begin{align}
H = H_{\text{R}} + H_{\text{L}} + H_{\text{T}}.
\end{align}

\enni For $s\tau_{3}-N-s\tau_{3}$ junctions,  $H_{\text{L/R}}$ are those of Eq.~\ref{Eq:BdG_st3_ld}, while for $s\tau_{3}-N-s$ junctions, $H_{\text{R}}$ is a single-channel $s$-wave bulk Hamiltonian, determined by the dispersion $\xi_{0}$ and $\Delta$ which also enter the expression for the two-orbital $H_{\text{L}}$. The two leads are connected via a tunneling Hamiltonian $H_{\text{T}}$. For the $s\tau_{3}-N-s\tau_{3}$ junctions, this takes the form 

\enni \begin{align}
H_{\text{T}, s\tau_{3}-N-s\tau_{3}} = & \sum_{\mathbf{k} \mathbf{p} \sigma} 
\bigg( T^{11}_{\mathbf{k} \mathbf{p}} c^{\dag}_{\text{R}, 1 \mathbf{k} \sigma} c_{\text{L}, 1 \mathbf{p} \sigma} 
+ T^{21}_{\mathbf{k} \mathbf{p}} c^{\dag}_{\text{R}, 2 \mathbf{k} \sigma} c_{ \text{L}, 1 \mathbf{p} \sigma}
\notag \\ 
+ & T^{12}_{\mathbf{k} \mathbf{p}} c^{\dag}_{ \text{R}, 1 \mathbf{k} \sigma} c_{ \text{L}, 2 \mathbf{p} \sigma} 
+ T^{22}_{\mathbf{k} \mathbf{p}} c^{\dag}_{ \text{L}, 2 \mathbf{k} \sigma} c_{ \text{L}, 2 \mathbf{p} \sigma} 
+ \text{H.c.} \bigg),
\end{align}
 
\enni where $T^{ij}_{\mathbf{k} \mathbf{p}}$ are tunneling matrix elements, and where we introduced R and L indices for the two leads in addition to the $1,2$ indices for the two orbitals. For the $s\tau_{3}-N-s$ junction $H_{\text{T}}$ has the simpler form 

\enni \begin{align}
H_{\text{T}, s\tau_{3}-N-s} = & \sum_{\mathbf{k} \mathbf{p} \sigma} 
\bigg( T^{(1)}_{\mathbf{k} \mathbf{p}} c^{\dag}_{\text{R}, \mathbf{k} \sigma} c_{\text{L}, 1 \mathbf{p} \sigma} 
+ T^{(2)}_{\mathbf{k} \mathbf{p}} c^{\dag}_{ \text{R}, \mathbf{k} \sigma} c_{ \text{L}, 2 \mathbf{p} \sigma} 
\notag \\
+ & \text{H.c.} \bigg),
\end{align}

\enni where the tunneling matrix elements are labeled by a single orbital index. We shall neglect the spin indices for simplicity.

The total currents out of the L lead read 

\enni \begin{align}
\dot{I}_{\text{Tot}, s\tau_{3}-N-s\tau_{3}} = & -i e \left( \dot{N}_{\text{L},1}+\dot{N}_{\text{L}, 2} \right)
\\
= & -ie 
\sum_{\textbf{kp}} \sum_{ij} \left( T^{ij}_{kp} c^{\dag}_{\text{R}, i \mathbf{k}} c_{\text{L}, j \mathbf{p}} - \text{H.c.} 
\right)
\end{align}

\enni \begin{align}
\dot{I}_{\text{Tot}, s\tau_{3}-N-s} = & -i e \left( \dot{N}_{\text{L},1}+\dot{N}_{\text{L}, 2} \right)
\\
= & -ie 
\sum_{\textbf{kp}} \sum_{i} \left( T^{(i)}_{kp} c^{\dag}_{\text{R}, \mathbf{k}} c_{\text{L}, j \mathbf{p}} - \text{H.c.} 
\right),
\end{align}

\enni where $N_{\text{L},i}$ is the total charge associated with either orbital in the L lead. 

As shown in Ref.~\cite{Mahan}, the Josephson current to leading order in the tunneling matrix elements is determined from

\enni \begin{align}
I_{\text{J}}(t) = & 2e \text{Im} \left[ e^{-2ieV t/\hbar} \theta(eU) \right],
\end{align} 

\enni where $eU$ is a potential drop across the junction. 

\subsection{$s\tau_{4}-N-s\tau_{3}$ and $s\tau_{3}-N-s$ junction}

For these cases the quantity $\theta$ is obtained via analytical continuation from 

\begin{widetext}

\enni \begin{align}
\theta_{s\tau_{3}-N-s\tau_{3}}(eU) = & \lim_{i \omega \rightarrow eU + i \eta}
\left[
 2 \sum_{ij} \sum_{mn} \sum_{\mathbf{kp} } 
T^{ij}_{\mathbf{kp}}  T^{mn}_{\mathbf{-k,-p}} \frac{1}{\beta} \sum_{i\nu} F^{\dag}_{\text{R}, im}(\mathbf{k}, i\nu) F_{\text{L}, jn}(\mathbf{p}, i\nu - i \omega)
\right]
\end{align}

\enni \begin{align}
\theta_{s\tau_{3}-N-s}(eU) = & \lim_{i \omega \rightarrow eU + i \eta}
\left[
 2 \sum_{ij} \sum_{\mathbf{kp} } 
T^{(i)}_{\mathbf{kp}}  T^{(j)}_{\mathbf{-k,-p}} \frac{1}{\beta} \sum_{i\nu} F^{\dag}_{\text{R}}(\mathbf{k}, \nu) F_{\text{L}, ij}(\mathbf{p}, i\nu - i \omega)
\right],
\end{align}

\end{widetext}

\enni where the anomalous Green's functions are determined from the corresponding L and R lead Hamiltonians as

\enni \begin{align}
F_{\text{L/R}, 11} (\mathbf{k}, i\nu) = &  \frac{- \Delta_{\text{L/R}}
\left[
(i\nu)^{2} - \left(\xi_{0} - \xi_{3}\right)^{2} - |\Delta_{\text{L/R}}|^{2} + \xi^{2}_{1} 
\right]
}
{
\Gamma(\mathbf{k}, i\nu)
}
\end{align}

\enni \begin{align}
F_{\text{L/R}, 12} (\mathbf{k}, i\nu) = &  \frac{- 2 \xi_{1} \Delta_{\text{L/R}}
\left(
i\nu + \xi_{3}
 \right)
}
{
\Gamma(\mathbf{k}, i\nu)
}
\end{align}

\enni \begin{align}
F_{\text{L/R}, 21} (\mathbf{k}, i\nu) 
= &  \frac{2 \xi_{1} \Delta_{\text{L/R}}
\left( 
i\nu - \xi_{3} 
\right)
}{
\Gamma(\mathbf{k}, i\nu)
}
\end{align}

\enni \begin{align}
F_{\text{L/R}, 22} (\mathbf{k}, i\nu) = & \frac{\Delta_{\text{L/R}} 
\left[
(i\nu)^{2} - \left(\xi_{0} +\xi_{3} \right)^{2}  - |\Delta_{\text{L/R}}|^{2} + \xi^{2}_{1}
\right]
}
{
\Gamma(\mathbf{k}, i\nu)
}
\end{align}

\enni for $s\tau_{3}$ leads, where 

\enni \begin{align}
\Gamma(\mathbf{k}, i\nu) = & \left( 
i\nu - E_{1}
\right)
\left( 
i\nu + E_{1}
\right)
\left( 
i\nu - E_{2}
\right)
\left( 
i\nu + E_{2}
\right)
\end{align}

\enni and 

\begin{align}
E_{1,2} = 
\sqrt{
\xi^{2}_{0} + \xi^{2}_{3}
+ \xi^{2}_{1} + |\Delta|^{2}
\pm 2 
\sqrt{\xi^{2}_{0} 
\left(\xi^{2}_{1} + \xi^{2}_{3}
\right)
+ \xi^{2}_{1} |\Delta|^{2}}
}
\end{align}

\enni are the BdG bands corresponding to the eigenvalues of the Hamiltonian in Eq.~\ref{Eq:BdG_st3_ld}.
Note that all $\xi$ and $E$ terms are functions of $\mathbf{k}$. Since we consider $|\Delta_{\text{L}}| = |\Delta_{\text{R}}|$,  $\Gamma$'s are independent of the lead index. 

For the single-channel $s$-wave lead, we have the standard expression 

\enni \begin{align}
F_{\text{R}} (\mathbf{k}, \nu) = &  \frac{- \Delta_{\text{R}}
}
{
\left[ 
i\nu - \sqrt{\xi^{2}_{0} + |\Delta|^{2}}
\right]
\left[ 
i\nu + \sqrt{\xi^{2}_{0} + |\Delta|^{2}}
\right]
}
\end{align}

These expressions simplify considerably if we consider the case where all of the tunneling coefficients are identical for all channels as in $T^{ij}_{\mathbf{kp}} = T_{\mathbf{kp}}$ and $T^{(i)}_{\mathbf{kp}} = T_{\mathbf{kp}}$. These correspond to junctions and the bound state spectra discussed in sections~\ref{Sec:st3} and~\ref{Sec:s_trvl}. The expressions for the $\theta$ functions are

\begin{widetext}

\enni \begin{align}
\theta_{s\tau_{3}-N-s\tau_{3}}(eU) = & \lim_{i \omega \rightarrow eU + i \eta}
 2 \sum_{\mathbf{kp} } 
T_{\mathbf{kp}}  T_{\mathbf{-k,-p}} 
 \frac{1}{\beta} \sum_{i\nu} 
\left[ \sum_{im}
 F^{\dag}_{\text{R}, im}(\mathbf{k}, i\nu) \right] 
 \left[ 
 \sum_{jn}
 F_{\text{L}, jn}(\mathbf{p}, i\nu - i \omega)
 \right]
 \label{Eq:Jsph_st3}
\end{align}

\enni \begin{align}
\theta_{s\tau_{3}-N-s}(eU) = & \lim_{i \omega \rightarrow eU + i \eta}
 2 \sum_{\mathbf{kp} } T_{\mathbf{kp}}  T_{\mathbf{-k,-p}}
 \frac{1}{\beta} \sum_{i\nu} F^{\dag}_{\text{R}}(\mathbf{k}, \nu) 
\left[ \sum_{ij}
 F_{\text{L}, ij}(\mathbf{p}, i\nu - i \omega)
 \right],
\end{align}

\end{widetext}

The sums in the square brakets amount to 

\enni \begin{align}
\left[ 
\sum_{ij}
 F_{\text{L/R}, ij}(\mathbf{k}, i\nu )
 \right] 
  = & 
\frac{
-4\Delta_{\text{L/R}} \xi_{3} \left( \xi_{0} + \xi_{1} \right)
}
{
\Gamma(\mathbf{k}, i \nu)
}
\label{Eq:Jsph_st3_sm}
\end{align}

\enni From this expression, it is clear that, in the limit of zero intra-orbital hybridization $\xi_{3}=0$, the \emph{total} Josephson currents in the tunneling limit for either $s\tau_{3}-N-s\tau_{3}$ or $s\tau_{3}-N-s$ junctions vanish identically. This holds even for a  finite inter-orbital hybridization $\xi_{1}$ on the L (and R lead if appropriate). The immediate reason for this remarkable result is that the L lead continues to preserve the orbital exchange symmetry defined in Sec.~\ref{Sec:Orbl_exch} \emph{in the bulk} which ensures that 

\enni \begin{align}
F_{\text{L}, 11}(\mathbf{k}, i\nu) = & - F_{\text{L}, 22}(\mathbf{k}, i\nu) 
\\
F_{\text{L}, 12}(\mathbf{k}, i\nu) = & - F_{\text{L}, 21}(\mathbf{k}, i\nu) 
\end{align}

\enni as long as the intra-orbital hybridization terms $\xi_{3}$ vanish, even though the complete Hamiltonian does not preserve the symmetry for the $s\tau_{3}-N-s$ junction. In the limit considered here, where all of the orbitals couple identically across the junction, this symmetry in the bulk of the L lead ensures that the total current out of the L lead vanishes.

Josephson currents which vanish in the limit of zero intra-orbital hybridization terms are completely consistent with the results of sections~\ref{Sec:st3} and~\ref{Sec:s_trvl} where the bound state spectra were shown to be insensitive to the relative phase precisely when the same intra-orbital hybridization terms were ignored.

\subsection{$s\tau_{0}-N-s\tau_{0}$ and $s\tau_{0}-N-s$ junction}

The expression for the current and the associated $\theta$ functions in Eqs.~\ref{Eq:Jsph_st3} is the same as for the $s\tau_{3}$ cases. However, the anomalous Green's functions for $s\tau_{0}$ are 

\enni \begin{align}
F_{\text{L/R}, 11} (\mathbf{k}, i\nu) = & \frac{\Delta 
\left[ 
-(i\omega)^{2} + \left(\xi_{0} - \xi_{3} \right)^{2}_{2} + \xi^{2}_{1} + |\Delta|^{2}
\right]
}
{\chi(\textbf{k}, i\nu) }
\end{align}

\enni \begin{align}
F_{\text{L/R}, 12} (\mathbf{k}, i\nu) = &   \frac{2 \Delta  \xi_{1} \xi_{0}
}{\chi(\textbf{k}, i\nu) } 
\end{align}

 \enni \begin{align}
F_{\text{L/R}, 21} (\mathbf{k}, i\nu) = & \frac{2 \Delta \xi_{1} \xi_{0}}
{\chi(\textbf{k}, i\nu) }
\end{align}

\enni \begin{align}
F_{\text{L/R}, 22} (\mathbf{k}, i\nu) = & \frac{\Delta
\left[
-(i\omega)^{2} + \epsilon^{2}_{1} + \xi^{2}_{1} + |\Delta|^{2}
\right]
}
{\chi(\textbf{k}, i\nu) }
\end{align}

\enni where 

\enni \begin{align}
\chi(\textbf{k}, i\nu) 
= \left( i\nu - E_{1} \right)
\left( i\nu + E_{1} \right)
\left( i\nu - E_{2} \right)
\left( i\nu - E_{2} \right)
\end{align}

\enni and  

\enni \begin{align}
E_{1/2}= &\sqrt{
\left( 
\xi_{0} \pm \sqrt{
\xi^{2}_{1} + \xi^{2}_{3} }
\right)^{2} + |\Delta|^{2}
}
\end{align}

\enni is the bulk BdG spectrum for $s\tau_{0}$ pairing. 

We calculate 

\enni \begin{align}
\left[ 
\sum_{ij}
 F_{\text{L/R}, ij}(\mathbf{k}, i\nu )
 \right] 
 = & \frac{
-2 \Delta 
\left\{ 
(i\omega)^{2} - 
\left[ 
(\xi_{0} + \xi_{1})^{2} + \xi^{2}_{3} + |\Delta|^{2} 
\right]
\right\}
}
{\chi(\textbf{k}, i\nu) }
\end{align}

\enni This is clearly distinct from the expression for $s\tau_{3}$ pairing in Eq.~\ref{Eq:Jsph_st3_sm}. Indeed, in the $\xi_{1}=\xi_{3}$ limit corresponding to degenerate orbitals this reduces to 

\enni \begin{align}
\left[ 
\sum_{ij}
 F_{\text{L/R}, ij}(\mathbf{k}, i\nu )
 \right] 
= & \frac{
-2 \Delta 
}
{
\left[
(i\omega
)^{2} 
- \xi^{2}_{0} - |\Delta|^{2}
\right]} 
\end{align}

\enni which amounts to twice the result for a single-channel junction, due to the presence of the two, decoupled orbitals. This is consistent with the dependence of the bound state spectrum on $\phi$ (Appendix~\ref{App:st0}).

\end{document}